\documentclass[epsfig,usenatbib,usegraphicx,]{mn2e}
\usepackage{tabularx}

\begin{document}

\title[The Luminosity Function in Galaxy Groups]{The Variation of the Galaxy Luminosity Function with Group Properties}
\author[A. Robotham et al.]
{Aaron Robotham$^{1,2}$\thanks{asgr@st-andrews.ac.uk},
Steven Phillipps$^{1}$\thanks{s.phillipps@bristol.ac.uk} and  
Roberto De Propris$^{3}$\\
$^{1}$Astrophysics Group, H.H. Wills Physics Laboratory, University of Bristol, Tyndall Avenue,  Bristol BS8 1TL, UK\\
$^{2}$School of Physics and Astronomy, University of St. Andrews, North Haugh, St. Andrews, Fife KY16 9SS, UK\\
$^{3}$Cerro Tololo Inter-American Observatory, Casilla 603, La Serena, Chile}

\date{Zeroth Draft. Accepted YYYY Month DD. Received YYYY Month DD; in original form YYYY Month DD}

\pagerange{\pageref{firstpage}--\pageref{lastpage}} \pubyear{YYYY}

\maketitle

\label{firstpage}

\begin{abstract}
We explore the shape of the galaxy luminosity function (LF) in groups of different mass by creating composite LFs over large numbers of groups. Following previous work using total group luminosity as the mass indicator, here we split our groups by multiplicity and by estimated virial (group halo) mass,
and consider red (passive) and blue (star forming) galaxies separately.
In addition we utilise two different group catalogues (2PIGG and Yang et al.) in
order to ascertain the impact of the specific grouping algorithm and further investigate the environmental effects via variations in the LF with position in groups. Our main results are that LFs show a steepening faint end for early type galaxies as a function of group mass/ multiplicity, with a much suppressed trend (evident only in high mass groups) for late type galaxies. Variations between LFs as a function of group mass are robust irrespective of which grouping catalogue is used, and broadly speaking what method for determining group `mass' is used. We find in particular that there is a significant deficit of low-mass passive galaxies in low multiplicity groups, as seen in high redshift clusters. Further to this, the variation in the LF appears to only occur in the central regions of systems, and in fact seems to be most strongly dependent on the position in the group relative to the virial radius. Finally, distance-rank magnitude relations were considered. Only the Yang groups demonstrated any evidence of a correlation between a galaxy's position relative to the brightest group member and its luminosity. 2PIGG possessed no such gradient, the conclusion being the FOF algorithm suppresses the signal for weak luminosity--position trends and the Yang grouping algorithm naturally enhances it.

\end{abstract}


\section{Introduction}
It is well known that galaxy properties, such as their morphology, star formation rate and colour, vary with the density of their local environment \citep{dres80, lewi02, kauf04}. Unsurprisingly, this is also seen as a variation of the galaxy luminosity function (LF) with
environment \citep[e.g.][]{ferg91, driv98, phil98, zabl00, tren02, hogg03, prac05, crot05}. These variations are likely due to some combination of the suppression of star formation once a threshold local density is reached \citep[e.g.][]{balo04, tana05} and the significant differences in the interaction and/or merger histories between environments \citep[e.g.][]{moor98, bell06}. 

In a previous paper \citep{robo06} we explored the variation of galaxy populations across the mass range from small groups up to clusters, in terms
of the LF parameters for `composite' groups. Using groups from the 2PIGG catalogue \citep{eke04a}, we showed that the characteristic magnitude $M^*$ brightened and the faint end slope $\alpha$ steepened with increasing group luminosity. That is, assuming that the total luminosity of the group members is a reasonable proxy for the total mass of the (group or cluster sized) halo containing them, more massive halos contain both brighter galaxies and larger fractions of faint galaxies. This, and the variations seen separately in the blue and red galaxy sub-populations, appeared consistent with a generic picture of the build up of the giant cluster galaxies through mergers and the amplification of the dwarf end of the LF as `field' irregulars `fall in' to larger groups and cluster and are transformed, through truncation of star formation, into dwarf ellipticals. Most of this transformation seemed to take place in quite low luminosity groups, i.e.\ quite early in the hierarchical build-up of the larger systems \citep[e.g.][]{hash00,wilm05}.

In the current paper, we extend this work \citep[and that of][]{eke04b}
by using different LF estimators, several different ways of dividing the groups by `size' and two different group catalogues. In addition we explore the `environment' in more detail by exploring radial LF variations within groups/clusters .

Section 2 describes the input group catalogues, noting in particular the
differences in grouping algorithms and the influence this may have on the
subsequent results. Section 3 discusses the various (group halo) mass proxies and derives the LF parameters for each of our composite groups obtained from the 2PIGG catalogue. Section 4 presents the equivalent discussion for groups in the catalogue of \citet{yang05b}, Section 5 considers the question of the positional dependence of galaxy luminosities within groups and Section 6 summarises our results.

We assume a standard cosmology with $\Omega_M = 0.3$, $\Omega_{\Lambda} = 0.7$ and a Hubble constant $H_0 = 100h\,$km$\,$s$^{-1}$. 

\section{Galaxy Groups and Group Catalogues}

Since this work is based on the group/cluster environment, it will be useful to quantify the defining characteristics of these systems. We define the virial radius to be the radius within which the average group/cluster mass (dark and luminous) density is 200 times the critical density of the universe. These radii typically fall in the range $0.3h^{-1}$Mpc\ $\leq r_{200} \leq 2h^{-1}$Mpc and the masses within them span $10^{12} M_{\odot} \leq M_{200} \leq 10^{16} M_{\odot}$ \citep{eke04a}. Of course, at the largest and most massive end of the scale these systems are typically called clusters, but the cut-off between the two classes of environment is largely arbitrary.

As well as affecting the evolution of their constituent galaxies, by playing host to the majority of galaxy-galaxy interactions \citep{cons05}, groups themselves undergo a large amount of dynamical evolution. In particular, the virial state of the group varies as a function of time; the calculated virial mass can vary hugely compared to the {\it true} mass over different dynamic time scales (\citet{mamo07} and references therein). Significantly, unless the group happens to be at the extreme of an expansion after an early collapse, the virial mass usually significantly {\em underestimates} the true mass for groups of all richness classes. In general, groups with a smaller crossing time to age ratio will possess virial and actual masses that are closer in value.
 
Groups are also key in larger scale hierarchical evolution as they merge frequently and rapidly in comparison to larger structures, and are the building blocks for cluster environments. Such a process of group build-up is a generic prediction of $\Lambda$CDM (\citet{lace93} {\it et seq.}), and it is found that merger rates are roughly proportional to the mass of the system, with the consequence that present day groups can trace their half-mass progenitor groups to a higher redshift than can clusters \citep{van-02}.

\subsection{Approaches to Grouping Galaxies}

The main trade-off that in general has to be made when choosing a grouping algorithm is whether it is more desirable to find all the bound members of a group and suffer some interlopers as a consequence, or to minimise the interloper rate and not group all bound objects. Tied into this compromise is the fact there is no {\it correct} answer when determining the strict boundaries of a group; depending on the over-density limit applied, the final product will have differing degrees of the infall region included in the final system. Many different approaches to grouping galaxy data are used \citep[see][for a recent discussion]{gerk05}, but the two most relevant ones for our work are discussed below.

\subsubsection{Friend-Of-Friend (FOF)}

All forms of the FOF algorithm rely in some way on taking a galaxy in the sample and searching the surrounding area/volume for other galaxies that meet certain criteria for separation in both velocity and projected coordinate space. The pioneering application of this process was by \citet{huch82}, where the catalogue of \citet{de-v75}, the earliest reasonably complete attempt at a group catalogue, was reconstructed using fully quantitative means. The 2PIGG catalogue used for this work is an adapted FoF algorithm which requires that linking criteria are met in both projection and radial separation. A brief overview of the 2PIGG algorithm is given below, but the interested reader is referred to \citet{eke04a} for a detailed description of the process.

When linking is considered, it is always in respect to how numerous objects are expected to be at a given redshift viz.

\begin{equation}
n(z,\theta)=\int_{b_{j}^{bright}}^{b_{j}^{lim}(\theta)}\phi[M(b_{j},z)]c_{z}(b_{j},\theta)d b_{j}
\end{equation}

\noindent where $\phi[M(b_{j},z)]$ refers to the Schechter luminosity function \citep{sche76} and has been determined for 2dFGRS in multiple papers; the function used for 2PIGG is that of \citet{norb02}. $c_{z}$ is a completeness correction determined from the position of a galaxy on a pixelated completeness mask, and typically scales the local density by a factor $\sim 0.8$. The integral is from the bright limit of the galaxy sample (effectively $b_{j}^{bright}=-\infty$) and the faint limit for each galaxy ($b_{j}^{lim}=19.4$ is the target magnitude for 2dFGRS, but the exact depth varies). $M(b_{j},z)$ uses the redshift of each galaxy, and hence luminosity distance, to convert the apparent magnitude into the absolute magnitude required for the LF. The next step is to calculate linking lengths for each galaxy that represent a certain degree of over-density relative to $n(z,\theta)$:

\begin{equation}
l_{\perp} = \mathrm{min} \left [ L_{\perp \mathrm{max}} (1+ z) , \frac{b}{n^{1/3}}\right]; l_{\parallel} = R l_{\perp},
\end{equation}

\noindent where $b$ is the scaling factor required for each galaxy link in order to achieve a certain degree of overdensity. Typically, N-body simulations opt for a relative over-density 125, so that the scaled linking length is b=0.2 times the mean interparticle separation (since the overdensity is $\propto b^{-3}$). This does not imply the mean galaxy separation factor $b_{gal}=0.2$ precisely, since dark matter and luminous galaxies (whilst tracing each other well in general) are not identically distributed. In fact it is predicted by \citet{cole00} that light is more concentrated than matter in a galaxy group halo, so $b_{gal}$ is likely to be smaller than 0.2 to achieve the same level of dark matter clustering. $R$ is the radial scaling desired, and must be larger than 1 in order to account for peculiar velocities within group haloes. Thus two galaxies $i$ and $j$, at a comoving distance $d_{c,i}$ and $d_{c,j}$ with angular separation $\theta_{ij}$, are linked if

\begin{equation}
\theta_{i,j} \leq \frac{1}{2} \left( \frac{l_{\perp,i}}{d_{c,i}} + \frac{l_{\perp,j}}{d_{c,j}} \right)
\end{equation}

and

\begin{equation}
|d_{c,i}-d_{c,j}| \leq \frac{l_{\parallel,i} + l_{\parallel,j}}{2}
\end{equation}

\noindent Neither $b$ nor $R$ are fixed, instead they are allowed to vary to account for the finding that small groups tend to have overestimated sizes and velocity dispersions and vice-versa for large groups (i.e.\ $b_{gal}$ has a slight dependence on group mass). Thus

\begin{equation}
b=b_{gal}\left( \frac{\Delta}{\Delta_{fit}} \right)^{\epsilon_{b}}
\end{equation}

\noindent and

\begin{equation}
R=R_{gal}\left( \frac{\Delta}{\Delta_{fit}} \right)^{\epsilon_{R}}
\end{equation}

\noindent where $\Delta$ is the observed overdensity of a cylinder with a comoving projected radius of $1.5h^{-1}$Mpc and an aspect ratio of $R_{gal}$. Applications to mock catalogues determine appropriate values of the parameters to be $\Delta_{fit}=5$, $\epsilon_{b}=0.04$ and $\epsilon_{R}=0.16$, and most significantly $b_{gal}=0.13$, $R_{gal}=11$ and $L_{\perp \mathrm{max}}=2h^{-1}$Mpc \citep{eke04a}. An extra limitation was applied to the FOF so that no perpendicular linking length could exceed  $2h^{-1}$Mpc, and by implication $22h^{-1}$Mpc in radial coordinates, though in reality such an extreme length will never be the limiting factor for the radial distance.

The combination of these parameters, and the use of a cylindrical rather than spherically optimised linking volume, was found to have the desired effect of maximising the truly grouped objects associated with a dark matter halo (group) and, depending on the redshift ranges used, minimising the interloper rate \citep{eke04a}. This latter factor was the main influence in determining a reasonable redshift limit for our subset of the group catalogue, as will be discussed in more detail later.
 
The main issue with using a FOF method is that the overdensities are by definition local, in fact since the overdensities are based on the nearest neighbour they could not be calculated with less data. Background tests carried out using Monte Carlo simulations of galaxy distributions \citep{robo08} demonstrate that in a volume with exactly the desired grouping density, the multiplicity of the system has a stochastic effect on how effectively the FOF algorithm groups all the objects. So even though the average density in the volume is precisely what we require, individual galaxies can be missed due to positional noise and nearby (sub)systems can avoid association.

After running a large number of realisations (100,000 divided by the system multiplicity), it was clear that the lower multiplicity systems suffered the former problem most: systems with multiplicities 5, 10, 20, 50, 100, 200 and 1,000 achieved 84\%, 89\%, 91\%, 94\%, 96\% and 97\% grouping completeness when using $1.0 S_o$ as the FOF linking length, where $S_o$ is the mean interparticle distance {\em within} the overdensity. As for the latter problem of subgroups not associating, the linking length that, on average, will link all objects via a FOF routine (essentially the overall percolation length) actually increases slightly as a function of multiplicity, changing from 1.1 to $1.3 S_o$ over the same range of multiplicities. Thus it is clearly important to consider which volume averaged overdensity is of interest when constructing a grouping algorithm, and whether it is worth increasing the interloper rate to ensure all truly associated objects are grouped; in the case of 2PIGG, the creators decided the answer to this question was yes. Thus FOF parameters were chosen so as to ensure nearly 100\% of true group members were successfully grouped, with the unavoidable side effect of a non-zero interloper rate.

\subsubsection{Halo Based Grouping and The Conditional Luminosity Function (CLF)}
 
Halo based methods for grouping centre on the premise that when creating a group catalogue the most important criterion to achieve is that all the grouped galaxies share a common dark matter halo, the early version of such a technique being that of \citet{post96}. The main principle behind such methods is that dark matter halos have a certain probabilistic distribution, and then depending on the mass of a particular halo the LF populating it is also probabilistic \citep[the Conditional Luminosity Function, CLF from here on; see][]{van-03}. Thus it seems intuitively clear that we can create a hypothetical Universe going one way -- building the halos then populating them with galaxies -- but what is required from a redshift survey is to take the galaxy data and try to go the other way and match them up with the expected distribution of dark matter halos. This is the approach used by \citet{yang05b} for grouping galaxies in the 2dFGRS and in the Sloan Digital Sky Survey (SDSS).

Their first step was to make a rough FOF catalogue of the survey in question using quite restrictive conditions: in the notation used in the previous section $b=0.05$ (rather than $\sim 0.13$) and $R=6$ (rather than $\sim 12$). The reason for such extreme overdensities was to find the likely group centres, not complete groups, and Yang et al. demonstrated this to be an effective means of locating the likely sites for halo centres. On top of the FOF centres, individual galaxies were also added to the list of potential centres if they were the brightest galaxies within a cylinder of radius $1 h^{-1}$Mpc and velocity depth $\pm 500$ kms$^{-1}$.

Once these potential centres were identified an iterative process of assigning galaxies to the halos was undertaken, and for a given halo-galaxy assignment calculation several variables were considered. The first was the current mass of the halo, as determined by the luminosity of all the galaxies grouped so far (taking into account luminosity dependent incompleteness) $L_{group}$. This was then extrapolated to find the total halo mass using:

\begin{equation}
L_{total}=L_{group}\frac{\int^{\infty}_{0}L \phi (L) dL}{\int^{\infty}_{L_{lim}}L \phi (L) dL},
\end{equation}
 
\noindent where $\phi$ is the LF (as above) and $L_{lim}$ is the minimum observable galaxy luminosity at the redshift of the group. The same LF function parameters were used throughout for this algorithm, but to return the expected CLF, Yang et al. used an empirically motivated luminosity dependent mass-to-light ratio. 
Interestingly though, they found that using a constant $M/L=400 h (M/L)_{\odot}$ returned almost identical groups, lending credence to the robust nature of their approach.

Based on these masses for the group halos, the group velocity dispersions and virial sizes could be estimated using the CDM halo density of \citet[][NFW hereafter]{navo96}

\begin{equation}
\rho(r)=\frac{\rho_{0}}{(r/r_{s})(1+r/r_{s})^{2}},
\end{equation}

\noindent where $r_{s}$ is the characteristic radius and $\rho_{o}$ is the density when $r/r_{s} \simeq 0.47$. Once dispersions and sizes had been estimated, any potential addition to the group was checked against the criterion that it was in a region of overdense phase-space relative to the group \citep[see][for details]{yang05b}.

When all possible additions to all of the groups had been made, the group properties were recalculated and the whole process repeated to see if the new groups could add any members that previously failed to meet the phase-space overdensity criteria. This algorithm was applied iteratively until the catalogue was stable and no new galaxies were grouped.
 
The major difference between the purely FOF approach to grouping and the slight mixture of techniques used for the halo based grouping is the scale at which a potential group member is considered. In the FOF algorithm purely galaxy-galaxy relations are calculated and by inference from this, groups of associated objects are constructed, leading to the possibility of incompleteness as discussed earlier. In the case of halo based grouping, the entire system built so far is used to ascertain grouping of any further galaxies; the only variables are the relative projected distance and radial velocity displacement from the currently constructed group. In comparison, the FOF approach does not vary the linking length as a function of what has already been grouped, thus a massive overdensity on one side of a galaxy does not extend the linking length, even though it intuitively should. The qualitative advantage of such an approach is that it requires less prior information in order to construct a catalogue; both approaches to group finding require knowledge of the LF, but only the halo based method requires knowledge of the density profile--- empirically well understood for clusters, but less so for small groups.
 
\subsection{Comparison of the 2PIGG and Yang-2dF Catalogues}

As discussed in detail above, the 2PIGG catalogue of \citet{eke04a} uses a FOF
algorithm, calibrated against a test mock galaxy survey, on galaxy data provided
by the 2dFGRS. Because of restrictions imposed by Eke et al. only 191,440 of the final 245,591 2dFGRS galaxies were used, though this is
significantly more than in the other three 2dFGRS based group catalogues of
\citet{yang05b} henceforth the Yang-2dF catalogue; \citet{merc02} which is not publicly available and \citet{tago06}. Of the known clusters in the volume concerned in the Abell \citep{abel89}, EDCC \citep{coll95} and APMCC \citep{dalt97} samples, the 2PIGG algorithm returns $51\%$ (84 of 166 clusters with a known redshift in SIMBAD) where the match distance in comoving space between identified group/ cluster centres must be within 2$h^{-1}$Mpc. The same match requirements for Yang groups only returns $28\%$ (47 of 166).

Compared to the other catalogues, 2PIGG has a large number of
groups where the number of members is $\geq 4$; 7,020 groups (comprised of 55,753 from 191,440 possible galaxies) meet this criterion, compared to 2,209 groups (from 60,000 galaxies) in Merchan \& Zandivarez, 2,471 groups (from 150,715 galaxies) in Yang-2dF, and 3,044 groups (from 184,495 galaxies) in Tago et al.
Proportionately 2PIGG is closest to the Merchan \& Zandivarez catalogue for this
figure. In terms of galaxies being grouped (2 or more members) or
not, 2PIGG is most similar to Yang-2dF, where $60\%$ are grouped
compared to 2PIGG's $55\%$. Both these figures are somewhat
different to the $30\%$ obtained by Tago et al.; Merchan \& Zandivarez only considered
groups of 4 or larger so cannot be used for this last comparison. 

A major difference between Yang-2dF (and also the Tago catalogue) compared to 2PIGG is their respective
treatments of group substructure; in particular the treatment of
single galaxies and halos. A single galaxy can be considered as
belonging to a small group, or alternatively it could be considered
to be part of the group halo as represented by just one galaxy. This problem of substructure was considered by \citet{burg04}, the conclusion being that most groups have substructure that
will merge over long dynamic times. The problem faced when
grouping is whether to count these merging groups as one group or as
many sub-groups. Yang et al. biased their algorithm towards more compact,
fully formed and gravitationally bound groups, hence the
significantly larger fraction of single galaxies in their results.
Both methods can be justified, but it is worth considering why the
results appear to be so different. To an extent it is a matter of
taste when considering this problem, but we should be aware later that the
results we obtain using the 2PIGG catalogue are based on groups
which may be in the merging process rather than fully formed (at
least relative to those in the Yang-2dF sample). The main effect of this will be realised
for low multiplicity groups, where the amount of clustering may seem
overemphasised.
 
\begin{figure*}[h]
\centerline{
    \mbox{\includegraphics[width=3.00in]{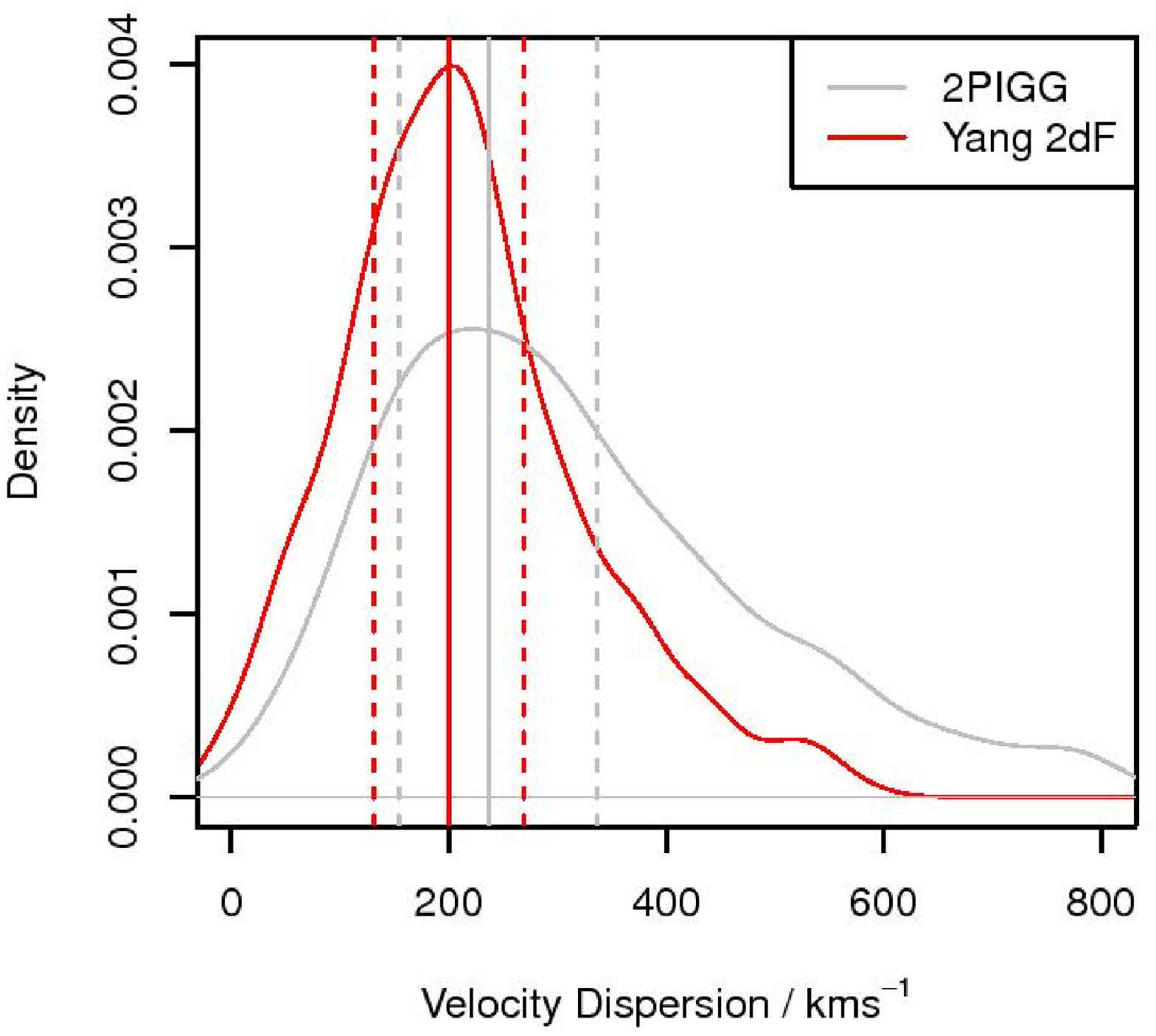}}
    \mbox{\includegraphics[width=3.00in]{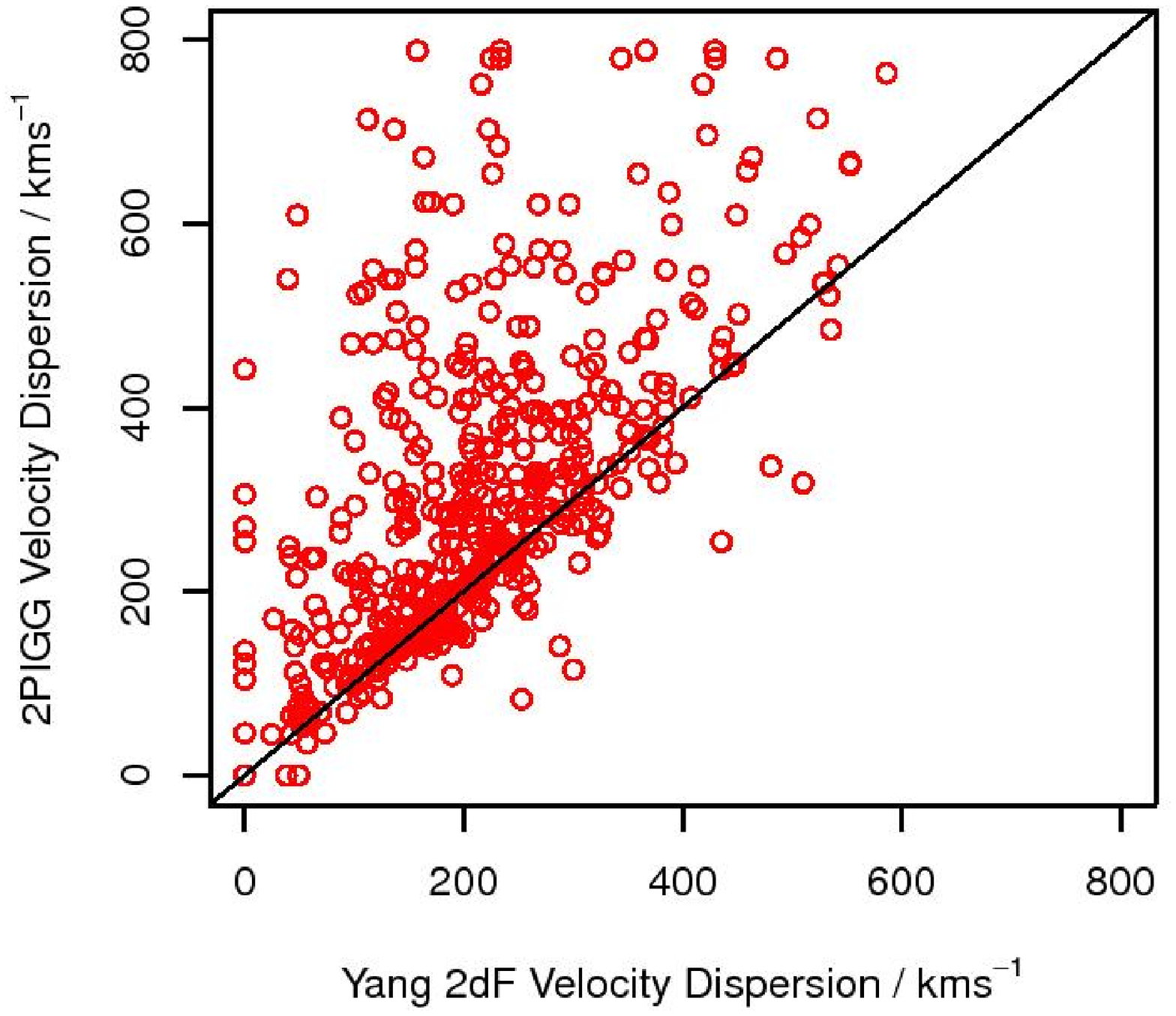}}
  }
 \caption{ \small Plots comparing the velocity dispersions for 2PIGG and Yang-2dF groups. The left panel is a comparison PDF for all systems with multiplicity larger than 4 and $0.05 \leq z \leq 0.1$. The solid vertical line indicates the median for each distribution, and the dotted lines are the inter-quartile range. The right panel shows the different values for the velocity dispersion between best matching groups. It is quite evident that in general 2PIGGs have greater velocity dispersions and a broader distribution of velocities. This is the case evident for matching systems also, so it is not merely an effect of the 2PIGG algorithm creating many large velocity dispersion groups that the Yang algorithm misses.}
  \label{veldisp}
\end{figure*}

\begin{figure*}[t]
  \centerline{
    \mbox{\includegraphics[width=3.00in]{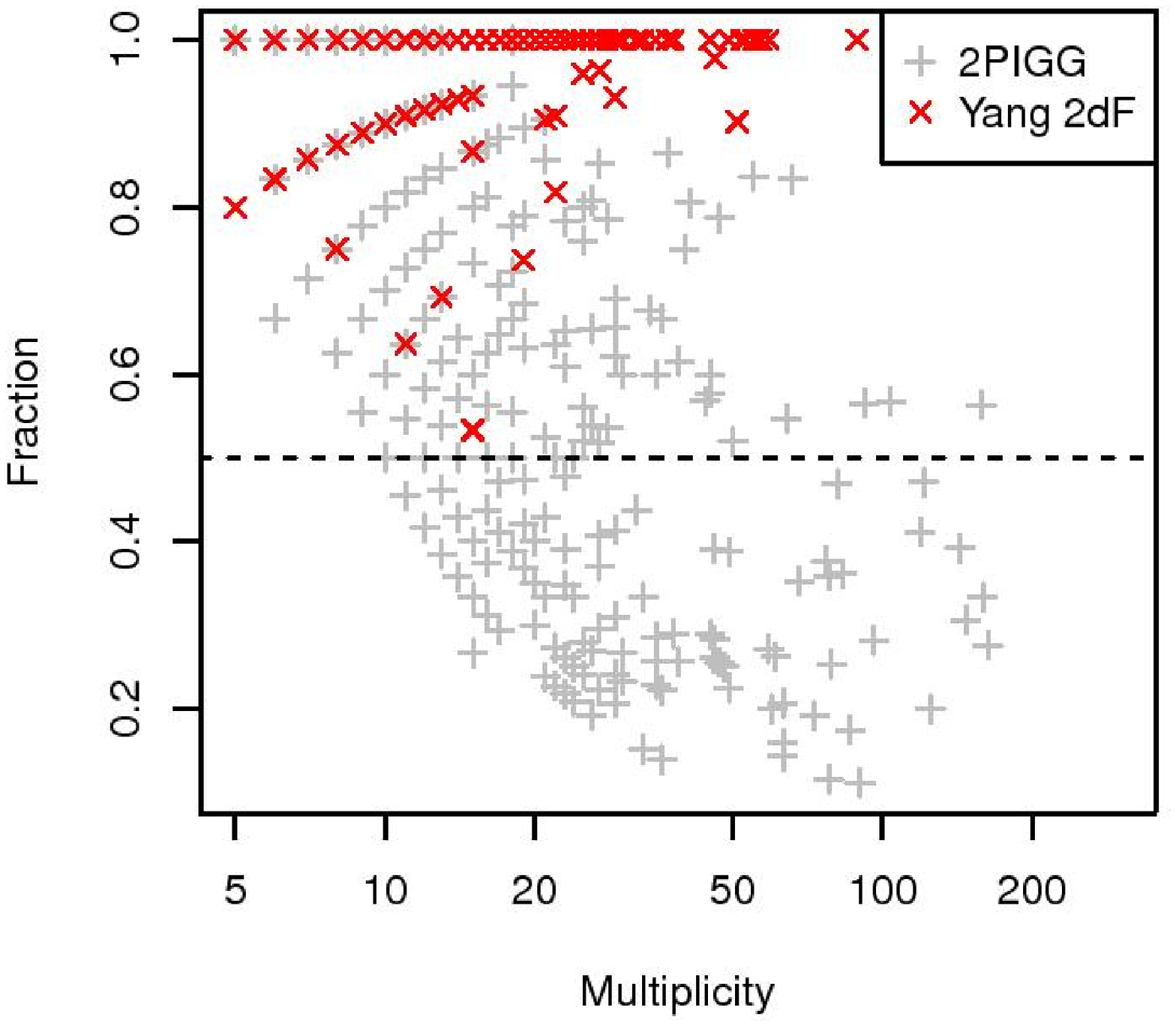}}
    \mbox{\includegraphics[width=3.00in]{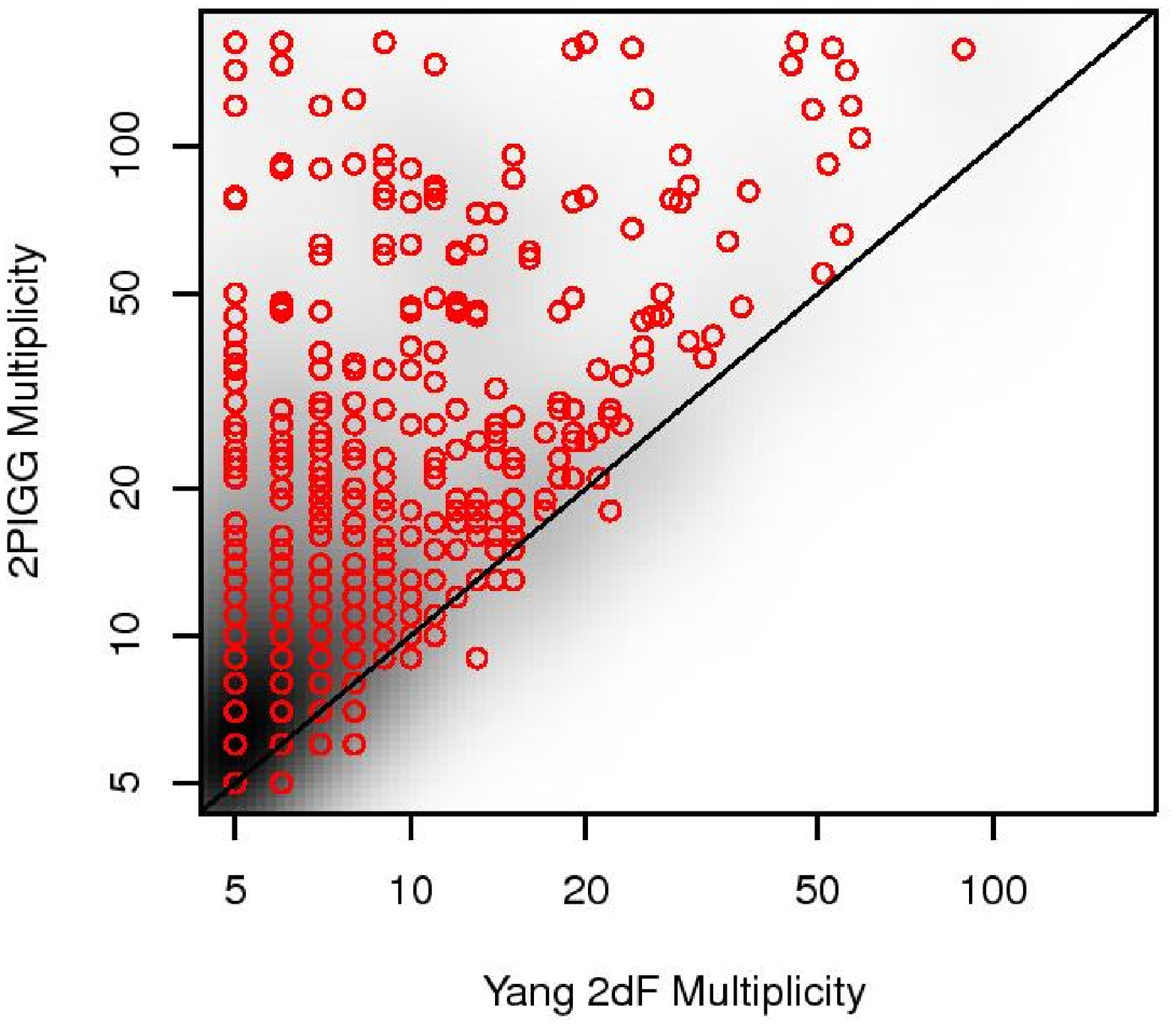}}
  }
\caption{ \small Left panel describes what fraction of each 2PIGG/ Yang group can be recovered in its best matching (in terms of two-way membership) Yang/ 2PIGG counterpart. A large number of Yang groups are 100\% contained within a 2PIGG, but the opposite is generally not the case. Right panel is a  comparison of group multiplicities for best matching group halos. The background grey-scale image is a guide to the local log-space density of objects, where darker regions reflect a larger number of groups.}
\label{hist}
\end{figure*}

Figure \ref{veldisp} contains two plots that demonstrate significant differences between the two group catalogues used in this work. For the full volume limited samples it is clear from the left panel that 2PIGGs have a higher typical velocity dispersion. The right panel compares velocity dispersions between groups that best match each other. The best match is defined as the two-way match that has the largest product of the galaxy matching fraction for each group. An example of matching groups would be a case where 4 galaxies are found in both a Yang-2dF group and a 2PIGG and the Yang-2dF group contains in total 5 galaxies and the 2PIGG contains 7. The product of these fraction is $\frac{4}{5} \frac{4}{7} \sim 0.46$. The best alternative matches might be, say, $\frac{1}{5} \frac{1}{1} = 0.2$ and $\frac{3}{7} \frac{3}{3} = 0.43$, so the match described first would be the preferential match. When reference is made to matched groups, it is in the sense described here. Obviously it is possible to devise alternative methods of group matching, e.g.\ based on spatial closeness or brightest galaxy, but for illustrative purposes the approach described is both simple to calculate and interpret. The left panel of Figure \ref{hist} demonstrates separately for Yang-2dF groups and 2PIGGs what fraction of each group is contained within its best matching counterpart. It is interesting to note that a much larger number of Yang-2dF groups are 100\% contained with the best matching 2PIGG. Also, a significant fraction of 2PIGGs share less than half their galaxies with the best matching Yang-2dF group, again indicating that on occasion multiple Yang groups are merged to form a single 2PIGG. 

Based on this matching, the right panel of Figure \ref{veldisp} displays an unambiguous tendency for 2PIGGs to possess much larger velocity dispersions for groups that are trying to describe the same halo. The median velocity dispersion of groups with multiplicity $\geq 5$ (calculated using the same GAPPER algorithm described in Eke et al. 2004a) is 199 kms$^{-1}$ for Yang-2dF groups and 238 kms$^{-1}$ for 2PIGG. In comparison, the best matching groups have median velocity dispersions of 206 kms$^{-1}$ for Yang-2dF groups and 282 kms$^{-1}$ for 2PIGG. This suggests that the groups created by 2PIGG that have no counterpart in the Yang-2dF catalogue are of lower velocity dispersion, since the median velocity dispersion is so much higher for matched 2PIGGs.

The most significant difference, in terms of how it might affect the present work, is the general tendency for the 2PIGG algorithm to link more objects compared to the Yang et al. group finding algorithm. Using the same redshift limits of $0.05\leq z \leq 0.1$ and a minimum of five galaxies in a group (see below), 2PIGG assigns 17,805 galaxies to 1,535 groups (average of 11.6 per group) compared to 5,364 galaxies in 544 groups (average of 9.9 per group) in Yang-2dF. Over the full multiplicity range, the Yang-2dF catalogue has a slightly higher proportion of low multiplicity groups (though less in total), and also has an upper multiplicity limit of 89 compared to 2PIGG's largest group of 163.

Inside the range considered in this work, 77 of the halos are found to be identical (comprised entirely of the same galaxies). In total 327 groups are unambiguous bijective matches, meaning more than 50\% of galaxy members in both the Yang group and 2PIGG can be found in the best two-way matching halo. Considering the Yang groups alone, 532 of the 544 groups considered have a 2PIGG system that contains more than 50\% of their assigned galaxies. This large fraction is possible because 47 2PIGGs contain multiple Yang groups that meet this criteria. The largest 2PIGG, a group comprised of 165 galaxies, contains 6 Yang groups that owe at least 50\% of their members to this single large 2PIGG.

In total we find 3,572 galaxies in the Yang-2dF groups from this sample of similar groups, compared to 6,383 galaxies in the 2PIGG groups. The right panel of Figure \ref{hist} compares the multiplicities between these similar groups found in the two different catalogues. There is a clear tendency for the 2PIGG catalogue to group additional galaxies compared to the Yang-2dF catalogue, in fact only 10 similar groups are larger in Yang-2dF. By the large quantity of bijective match halos we can be confident that they are, in general, describing the same dark matter halo, the main variance being the degree to which galaxies are found to be associated with it. We can conclude that the 2PIGG catalogue describes the same base population for a given halo as the Yang-2dF catalogue, but it also tends to include a more tenuously associated population, which may be less bound to the halo in question.

\section{Composite 2PIGG Luminosity Functions}

As input for our calculations we follow \citet{robo06} in first selecting all groups from the 2PIGG catalogue \citep{eke04a} which have $0.05 < z < 0.10$. The $z=0.05$ lower limit is motivated by the small 
volume sampled by 2PIGG at low redshift (so that groups may not be representative), 
while at $z=0.10$ the 2dFGRS apparent magnitude limit begins to exclude even moderate 
luminosity galaxies ($M_B < -18 + 5$log$\, h$) from the sample. [Note that hereafter the term 5 log$\, h$ will be suppressed for clarity, so unless otherwise indicated, absolute values will be quoted as if $h=1$.] In addition, over this range of redshift the interloper rates remain low and steady \citep{eke04a}.

Since 2PIGG includes groupings as small as 2 galaxies, we also impose a minimum group size
of 5 members. (For comparison, at the range of distances we use, the Local Group would appear to have a multiplicity of 2 to 4). This should increase the likelihood that the groups we select are 
actual bound physical systems. As discussed in \citet{robo06}, these choices leave 1535 
groups containing a total of 21,752 galaxies, when corrected for incompleteness. The incompleteness arises primarily as a result of fibre collisions in the 2dFGRS, that is two fibres have a minimum possible separation.  \citet{eke04b}
present details of the calculation and we follow their prescription, weighting each observed galaxy by a factor depending on its local galaxy density (as determined from the positions of its 10 nearest neighbours).  These groups can then be used in various combinations to determine composite luminosity functions
for galaxies in different environments, as described in the following subsections.

\begin{figure*}
 \centerline{
    \includegraphics[width=6in]{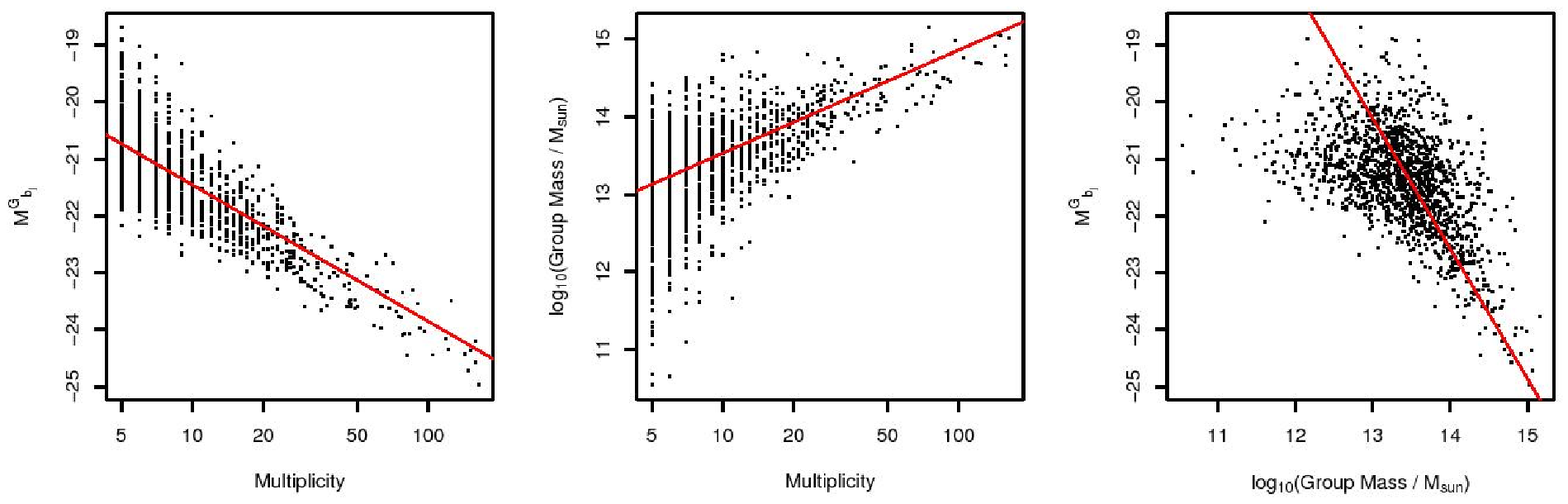}
  }
\caption{\small  Comparison of different methods for determining the relative {\it size} of galaxy groups. Red lines show the best fit linear trends in the data as a visual guide. The trend shown in right panel is fixed such that five magnitudes spans two decades in mass---as would be expected if the relationship between mass and luminosity scaled linearly.}
\label{sizecomp}
\end{figure*}

To derive the LFs, we first calculate absolute $M_{b_j}$ magnitudes and rest-frame $b_j-r_F$ colours 
for galaxies in the groups, following the prescriptions of \citet{cole05}. We ignore the relatively small (and uncertain) amount of evolution within the narrow redshift range we survey. In addition to the LFs derived for the entire sample we also create LFs for red (quiescent) and blue (star-forming) galaxies separately. For this we again follow \citet{cole05} and divide the samples at a colour of $b_j-r_F=1.07$.

Before attempting to determine any observed trends, we should consider the
way we quantify the group or cluster environment. A group's
total mass is evidently a critical measure, as halo mass is the key
variable in structure formation models (Press \& Schechter 1974 {\it et seq.}; 
Yang et al. 2005c). The total blue luminosity, $M^G_{b_j}$, as used in  \citet{robo06} has been demonstrated to be a good measure of (and therefore proxy for)
total stellar mass and total halo mass (Padilla et al. 2004, Yang et al. 2005d, 2005e),
but one could argue that 
the total $R$ band luminosity, $M^G_{r_F}$, would be preferable to the $B$ band luminosity, 
as it is less subject to contributions from short term star formation. In fact Robotham (2008) found essentially the same results when ordering groups by red luminosity as Robothem et al. (2006) did when using total blue luminosity.
We refer the reader to \citet{robo06} for details of our original results and for a comparison of our and earlier results on the variation of the LF with group luminosity (e.g. Eke et al. 2004b, Zandivarez, Martinez \& Merchan 2006).

Because different environmental indicators may each have their own merits, in the following, we use two different methods for sorting the data, a virial mass estimate, and the multiplicity (number of members). 
In general,these parameters all correlate with each other, and with group luminosity (see figure \ref{sizecomp}), but at a finer level they show disagreements as to which groups should be considered most similar.

\subsection{Method}

In \citet{robo06} we computed a composite galaxy LF, following the method described by \citet{coll89}.  To ensure the veracity of our results, for the present paper, we instead use a maximum likelihood (ML) method to fit the LFs for the composite groups.

A key improvement when using the ML method is that the data is not binned, instead a probability density function is constructed based on the Schechter luminosity function itself. To do this correctly the integrated luminosity function is scaled to 1 for each observation, so for any given galaxy this scaling factor is determined by the integral between $-\infty$ and the absolute magnitude limit of the survey for that galaxy---this is easily found by taking the apparent magnitude limit of the survey and subtracting the k-corrected distance modulus for the galaxy. Once the correct scale has been calculated, the log of the density of the PDF at the magnitude of the galaxy is determined and then this whole process is repeated for every galaxy in the subset of interest. The sum of the logs is finally calculated, giving us the `log-likelihood', and then the Schechter parameters $\alpha$ and $M^*$ are repeatedly changed. The combination of parameters that best fit the data, or strictly speaking the fit that yields the `most likely' data, is that which maximises the log-likelihood. The confidence limits in log-likelihood space are much like those of $\chi ^2$ space except compacted by a factor of 2, i.e.\ for the error about one parameter (so leaving 1 degree of freedom) the $95\%$ confidence limit is found at the log-likelihood value 1.92 less than the maximum, instead of 3.84 as would be expected when integrating the $\chi ^2$ distribution.

An advantage of using the maximum likelihood technique here is that it does not require the completeness at the faint end to be corrected for on a group by group basis, instead it is taken into account for every galaxy by virtue of the magnitude limit of the survey which is used to scale the PDF. This is somewhat akin to a $1/V_{max}$ approach to fitting, but due to the galaxies not being binned, extreme distortions at the faint end are not likely to occur. Thus this approach, whilst agreeing excellently with the previous paper in like-for-like systems, is not prone to binning errors. 

To determine the quality of fit to the data, if we assume the correct PDF has been applied to the data, then it would be expected that the distribution of the integrals of the Schechter luminosity function between $-\infty$ and $M_{b_j}$ for each galaxy should be uniformly distributed. The degree of uniformity for this distribution can be determined in a number of ways, but here the KS-test has been used. The p-value returned from the test will give us an indication of the quality of fit. This means that as well as obtaining the best fit given an assumed Schechter model, we also have a means of quantifying whether the model is appropriate.

\subsection{Virial Mass}

\begin{figure*}
 \centerline{
    \mbox{\includegraphics[width=3.00in]{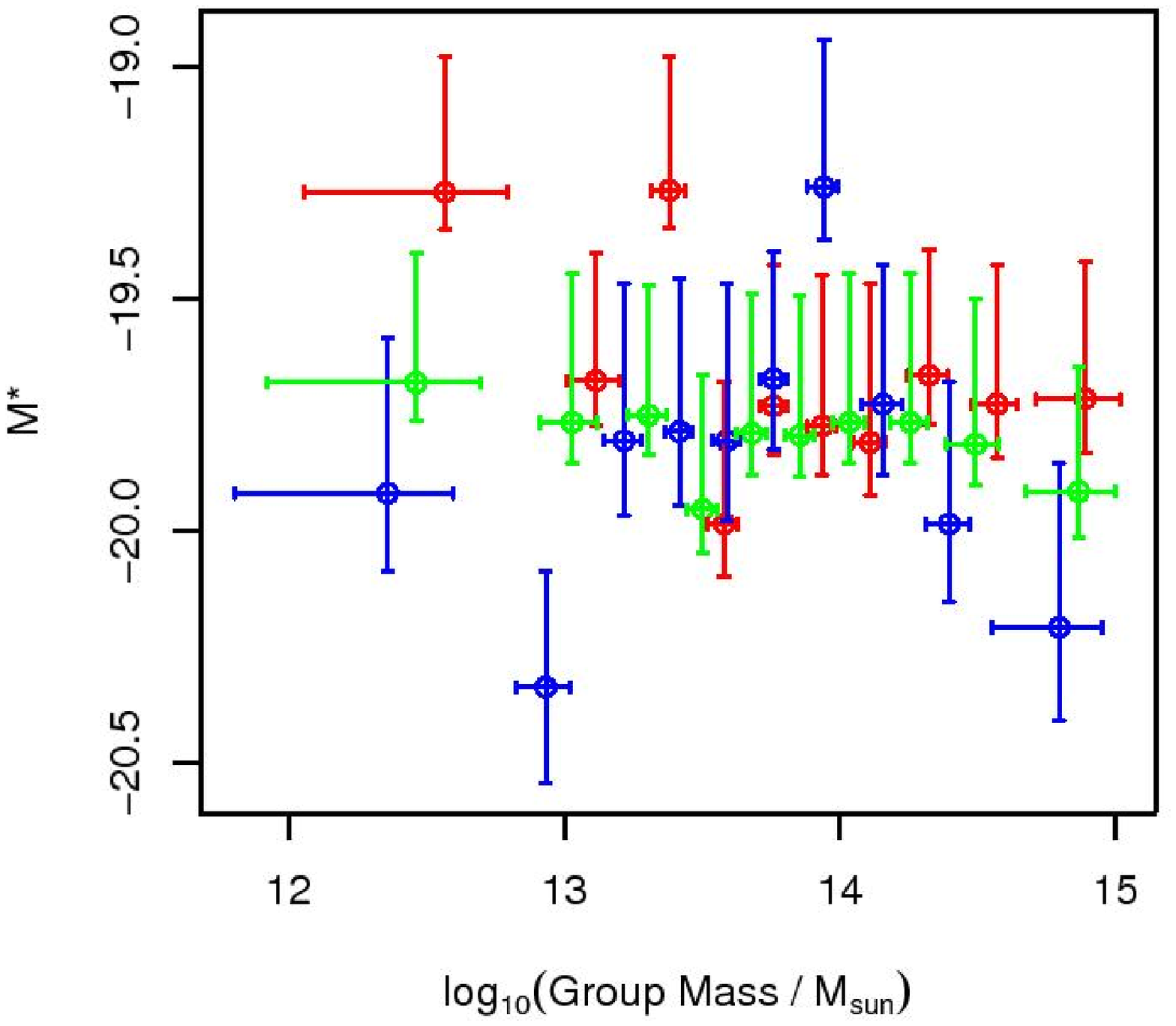}}
    \mbox{\includegraphics[width=3.00in]{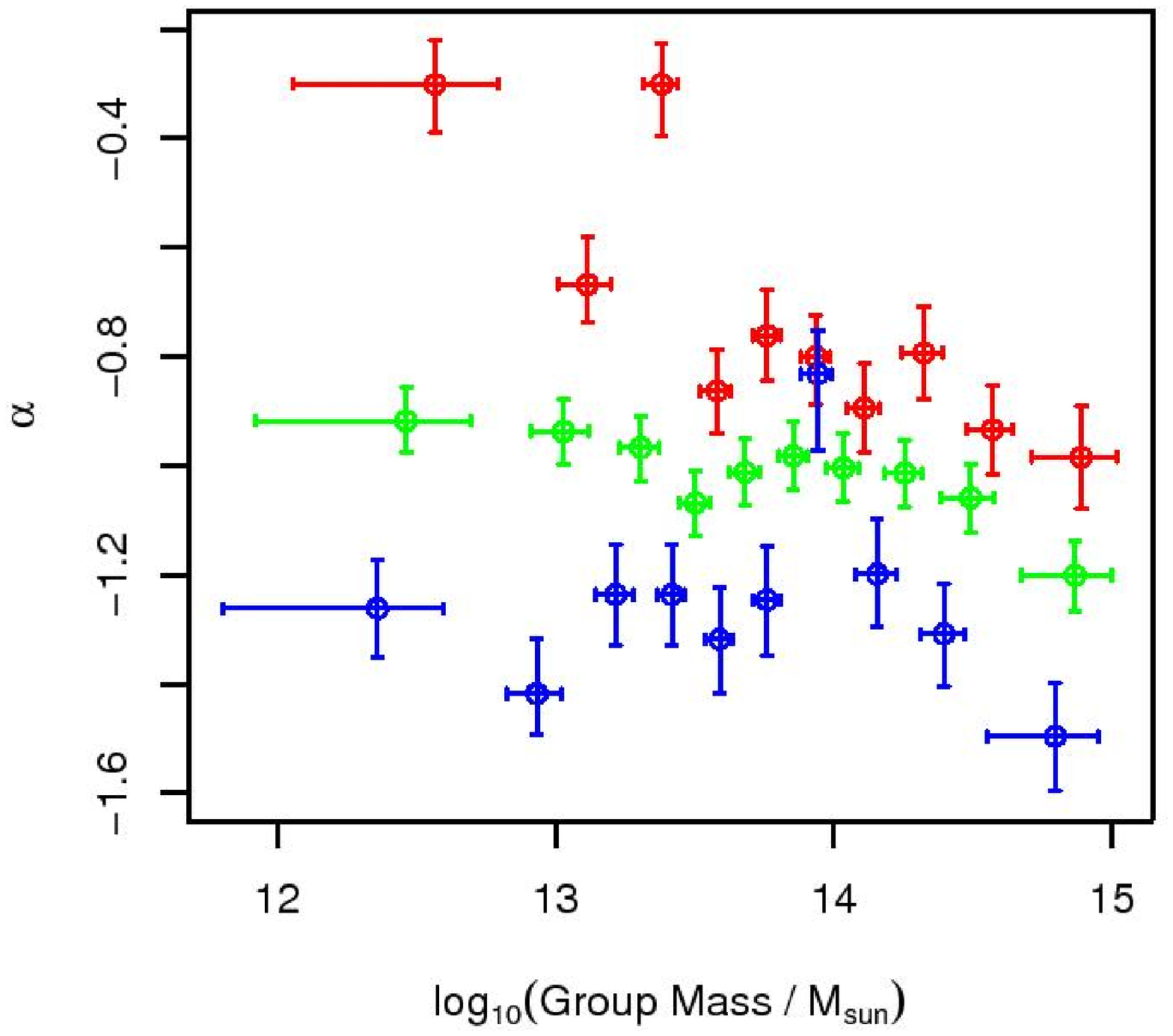}}
  }
 \caption{\small  Luminosity functions for groups from the 2PIGG catalogue as found via Maximum Likelihood for blue, red and all galaxies (blue, red and green colours respectively). Plots show best fit $M^*$ and $\alpha$ for groups sorted into 10 bins via virial mass estimates. x-axis errors are the standard deviation of the data, and the y-axis errors are the $95\%$ confidence levels computed from the log-likelihood.}
  \label{MaxLikeMass}
\end{figure*}

For the mass ordered data, we follow Robotham et al. (2006) and again use 10 bins with as near to equal numbers of galaxies as possible. Figure \ref{MaxLikeMass} contains the variations of $\alpha$ and $M^*$ as a function of virial mass as determined by the method of Eke et al. (2004b), using the radius (specifically the root mean square radial distance of galaxies from the group centre, $r_{rms}$) and velocity dispersion parameters included with the group catalogues. This assumes a degree of excess $KE$ compared to $GPE$ in the relative virialisation ($-2KE/GPE=1.2$), but was found to correspond well to the underlying group mass.

Both the total population and the red galaxy population show evidence of $M^*$ becoming brighter as a function of group mass, supporting the previous results, while the blue galaxy trend appears to be more complicated, with  $M^*$ first becoming fainter with group mass, and then brightening sharply for the most massive groups. However, for the sub-populations there is more scatter than with the luminosity orderings, so,
for instance, $M^*$ originally becoming fainter for the blue galaxies is quite possibly the result of an outlier at group mass $\sim 10^{14} M_{\odot}$ (given the suspiciously large $\alpha$ value at this point and the well known correlation of errors in $M^*$ and $\alpha$). The larger scatter is likely due, at least in part, to the fact that the calculated virial masses are not a particularly good indicator of mass until we reach the most massive groups \citep[see also][]{eke04b,mamo07}.
 Of course, in the virial mass case there is no (direct) selection against bright galaxies being in small, low mass groups (as there is for low luminosity groups), so any such cases might also increase the scatter. On the other hand the continued existence of the same trend, particularly for the red galaxies, confirms that these were not forced by the luminosity selection in the previous work.

The trend seen in $\alpha$ values is more straightforward, all galaxies combined show a subtle but recognisable pattern of $\alpha$ becoming steeper as a function of group mass, starting and finishing with similar values to the extremes seen in Robotham et al. (2006) and Robotham (2008): $-0.9$ for the least massive up to $-1.2$ for the most massive. When this is subdivided, the red galaxies show a very strong steepening trend as the faint end becomes systematically more populated as a function of group mass, whereas the blue galaxies are consistent with an almost constant $\alpha \sim -1.2$ \citep[c.f.][]{pope06,merc06}, with only the most massive systems displaying a notably steeper faint end ($\alpha \sim -1.5$), as frequently reported for clusters \citep[e.g.][]{driv94}. \citet{bell04} and \citet{cowi08} note a similar large change in the shape of the mass function of red galaxies in general, compared to that of blue galaxies, due to the steady migration of `blue cloud' galaxies to the `red sequence'. The early type LF thus tends to be affected at the faint end due to the recent creation of dEs \citep{smit08}, while once the late types ( of all luminosities) feel the affect of the dense environment they quickly cross the `green valley' and are removed from the late type sample measured here. This is, of course, consistent with the recent build-up of the red sequence after galaxies have fallen into groups and clusters (see, e.g., \citet{stot07,de-l07,smit09} for clusters, and \citet{tana05,hain08} for lower density environments).

The observed behaviour of the mass ordered groups confirms and strengthens the agreement with the predictions of the 
halo occupation model of \citet{yang05a} noted by Robotham et al. (2006). At low group mass (luminosity) Yang et al. find that the luminosity of the brightest 
galaxies (so, effectively, $M^*$) should increase quite steeply, with a much slower increase 
in the more massive groups, the 
turnover in slope occuring at halo masses a few times 
$10^{13} M_{\odot}$ \citep[see also][]{vale04,coor05}. \citet{yang05a} argue that this change in behaviour occurs at the transition from
efficient to inefficient cooling in different mass halos, and is the same
feature required in semi-analytical modelling of galaxy formation in order 
to match the bright end of the overall galaxy LF \citep[e.g.][]{bens03}, while Cooray \& Milosavljevic suggest that it marks the mass at which the dynamical friction timscale for orbital decay in the halo matches its age. Observationally,
Eke et al. (2005) find that the variation of group mass-to-light ratios (in the $K$ band) with group mass also levels off at this same point.

The changes in $M^{*}$ and $\alpha$ are also understandable physically in the context of group scale mergers. If groups hierarchically merged but there was no change in either $\alpha$ or $M^*$ then this would be a sign that the group merger had not prompted galaxy scale effects.

\subsection{Multiplicity}

For the multiplicity ordered data, we will again use 10 bins with as near to equal numbers of galaxies as possible, but a 3 bin version is also used to aid later analysis of results. For these subsets the sample is split into groups of fixed multiplicity ranges: $5\leq N\leq 9$, $10\leq N\leq 19$ and $N\geq 20$. These 3 bins contain 6516, 4393 and 6896 galaxies respectively.

\begin{figure*}
 \centerline{
    \mbox{\includegraphics[width=3.00in]{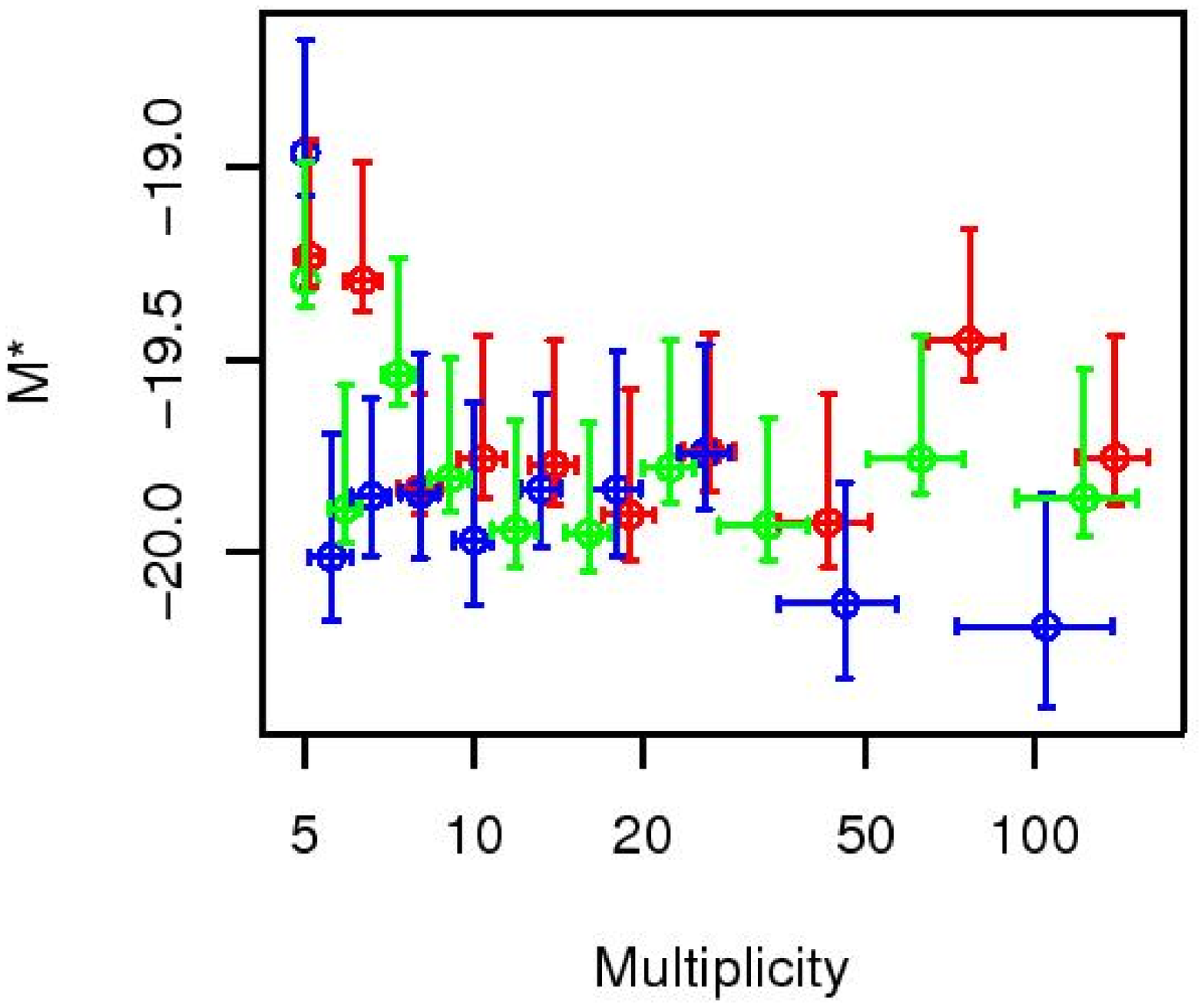}}
    \mbox{\includegraphics[width=3.00in]{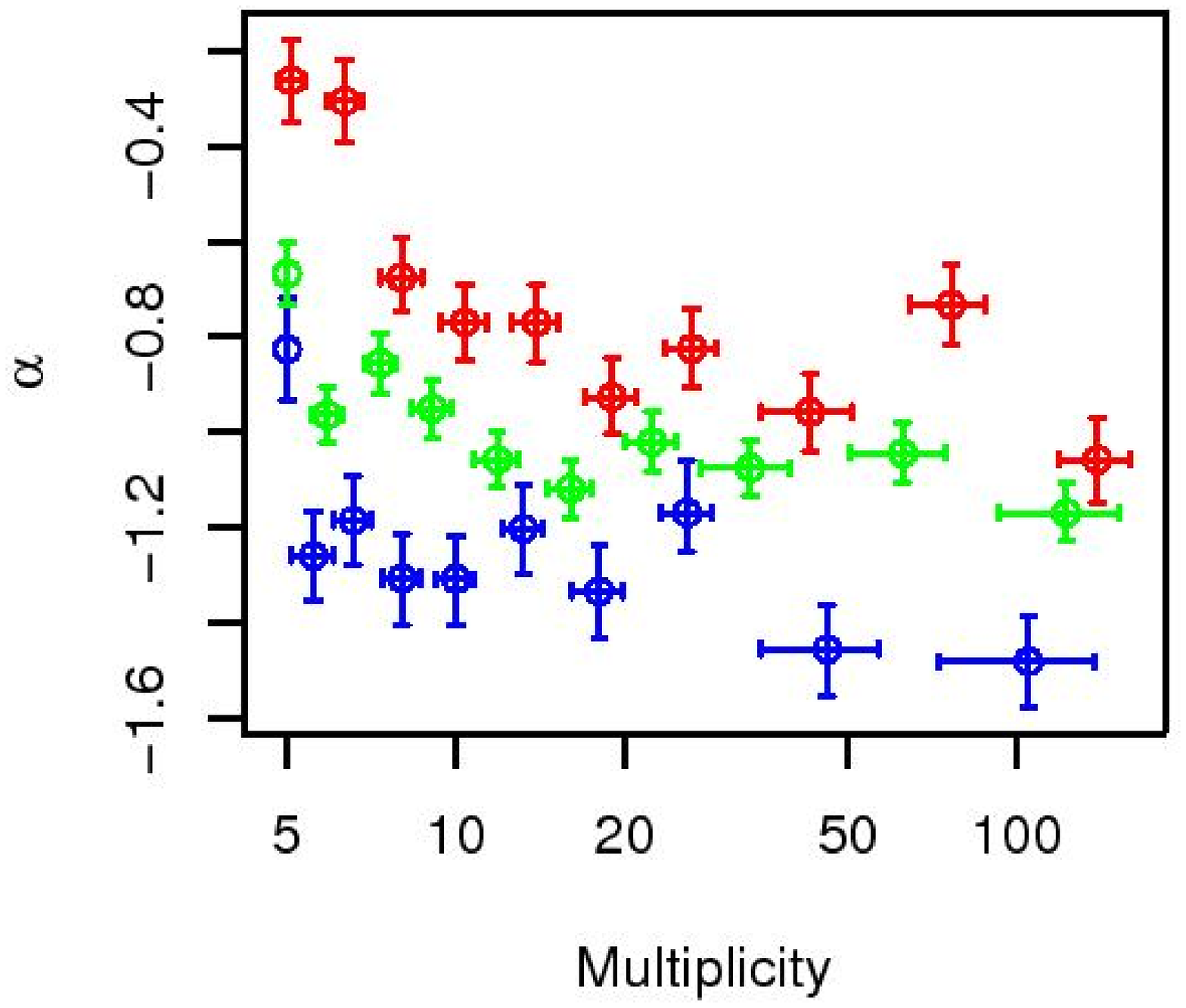}}
  }
   \centerline{
    \mbox{\includegraphics[width=3.00in]{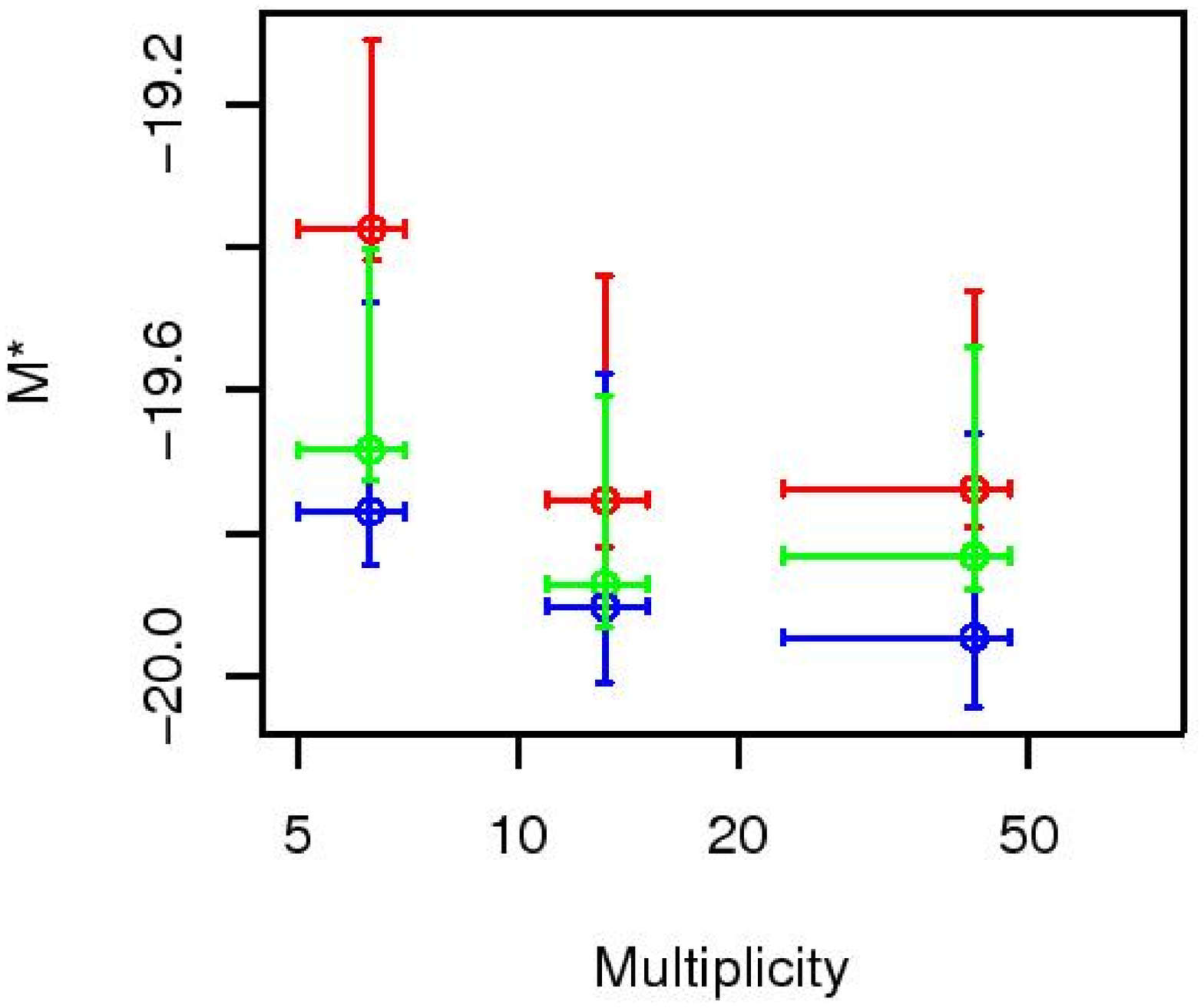}}
    \mbox{\includegraphics[width=3.00in]{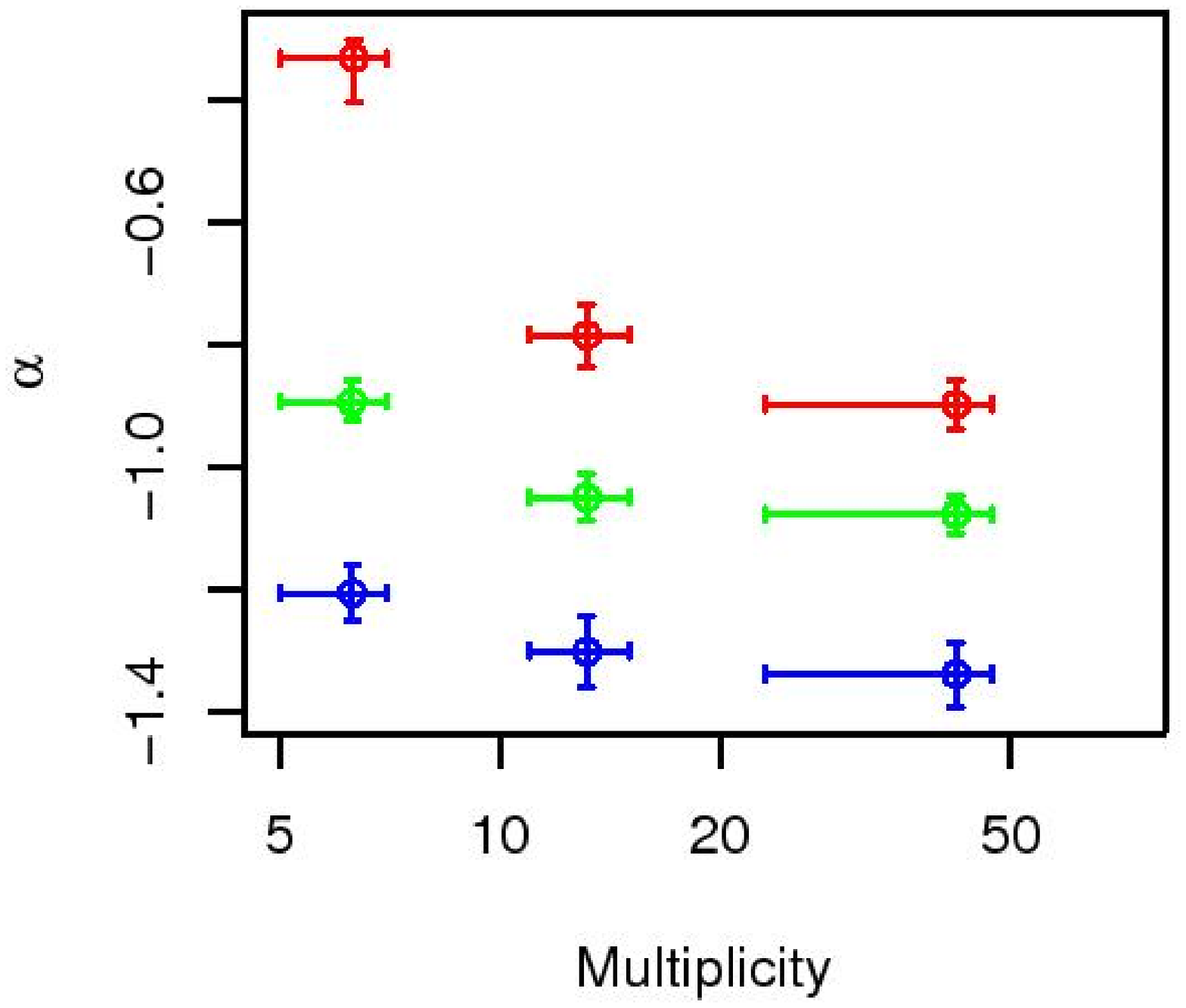}}
  }
  \caption{\small  Luminosity functions for groups from the 2PIGG catalogue as found via Maximum Likelihood for blue, red and all galaxies (blue, red and green colours respectively). Plots show best fit $M^*$ and $\alpha$. Both top and bottom plots are sorted by group multiplicity. The top plot has 10 splits based on maintaining galaxy count in bins. The bottom plot has three bins of roughly similar galaxy count, but specifically the 3 ranges: $5 \leq N \leq 9$, $10 \leq N \leq 19$ and $N\geq20$. x-axis errors are the $25\%$ and $75\%$ quantiles, and the y-axis errors are the $95\%$ confidence levels computed from the log-likelihood.}
  \label{MaxLikeMult}
\end{figure*}

 The results can be found in figure \ref{MaxLikeMult}. In terms of $M^*$, we see little systematic change for the overall population until we reach the very lowest multiplicities. Indeed, splitting by colour, the blue galaxies only have a fainter $M^*$ in the lowest bin. Generally, though, the trends are in line with our other group orderings. 

Interestingly, the lowest multiplicity systems, in this case all groups with only 5 members, also experience more dwarf suppression for blue galaxies than any of the other methods for sorting the data, with $\alpha\sim-0.8$. This implies that a system being numerically poor has more of an effect on the lack of late-type dwarfs than the mass of the system. The red galaxies have the same extreme value of $\alpha \sim -0.3$ as the small groups in the other orderings. At the other end, the $\alpha$ parameters settle to their normal rich system values, which is understandable since multiplicity correlates very tightly with virial mass in the largest systems (central plot of figure \ref{sizecomp}). 

In the lower panels of figure \ref{MaxLikeMult}, the data is split into just 3 bins, with specific multiplicity boundaries (at 10 and 20). The smoothing out of values is what you might expect when combining composite groups. The trend for $M^{*}$ is still not prominent for these larger composites, despite the smaller error bars, it being slightly fainter only in the $5 \leq N \leq 9$ bin. Even the early type galaxies have a range of only $M^{*}\sim-19.4$ to $M^{*}\sim-19.8$), only just outside of the $95\%$ confidence intervals, a much smaller effect than seen in mass ordered groups.

The trends in $\alpha$, on the other hand, are extremely similar to those seen for the other group orderings.
The differences in the best fit $\alpha$ values for early-type galaxies are again quite pronounced, and are clearly seen in the lower panel with only three multiplicity ranges.

\begin{figure*}
\centerline{
\mbox{\includegraphics[width=3in]{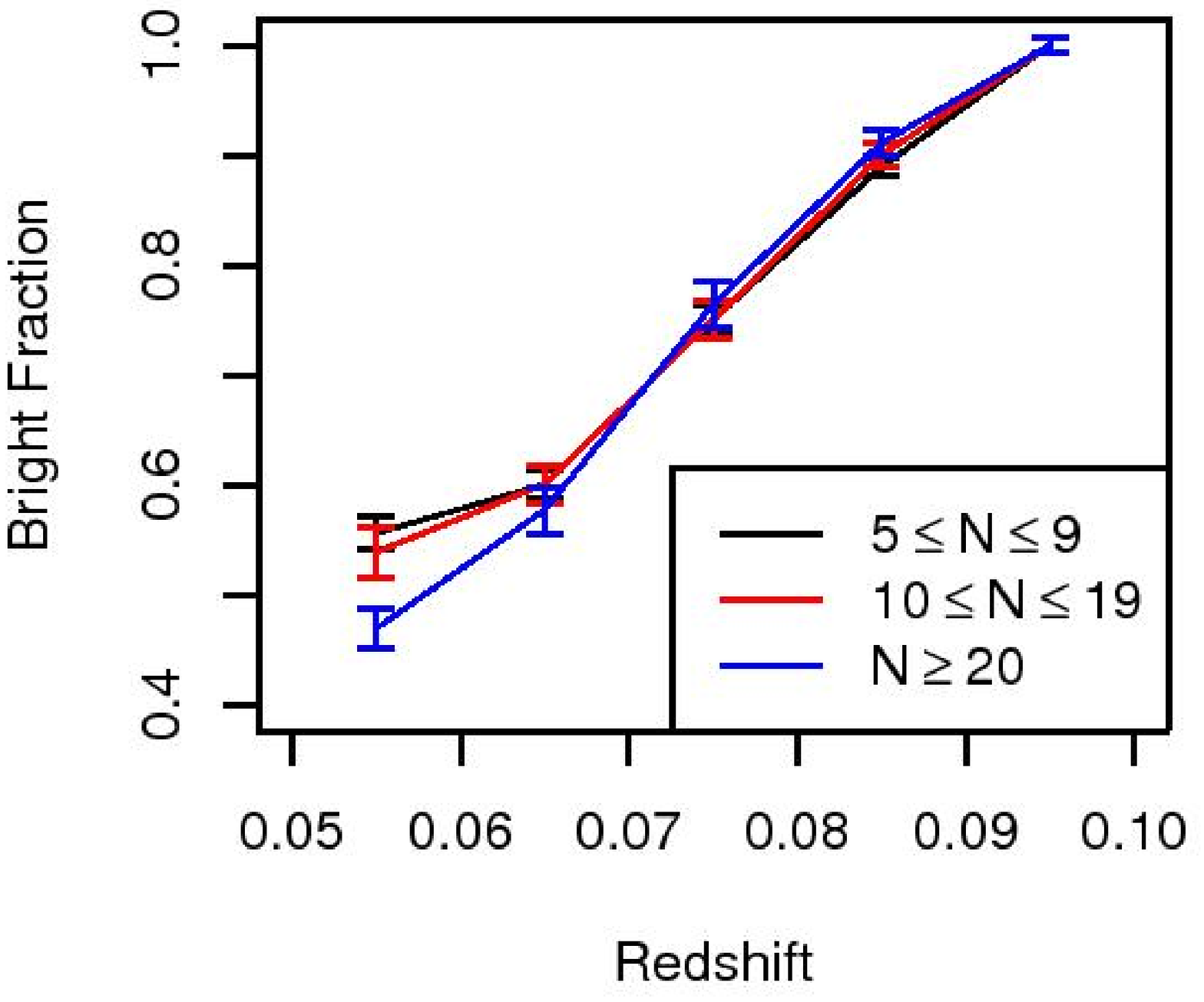}}
\mbox{\includegraphics[width=3in]{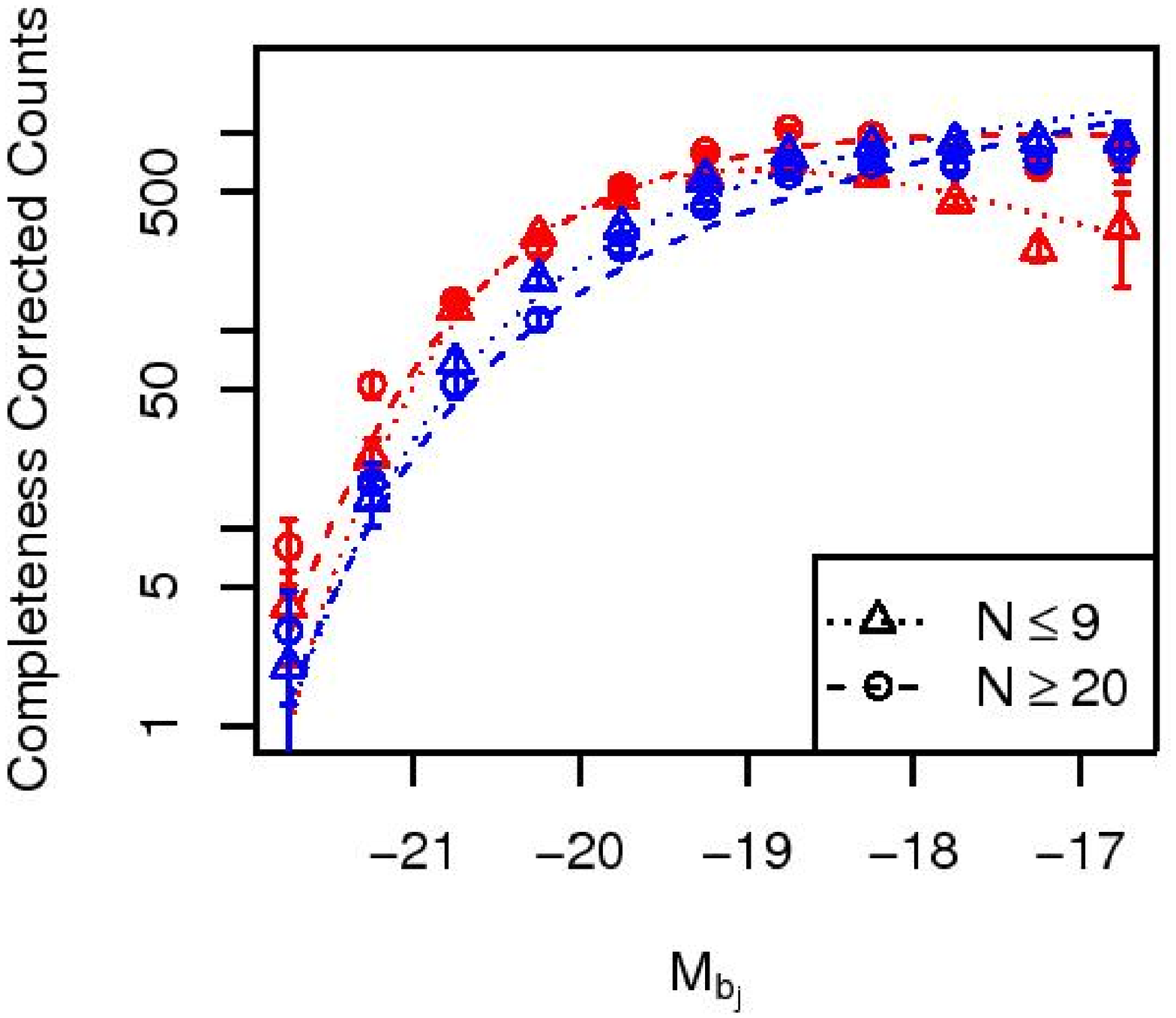}}
}
\caption{\small  Left plot shows the median fraction of galaxies with $M_{b_{j}}\leq -18.25$ in groups as a function of original multiplicity and redshift. It is quite clear that the lower multiplicity systems have a larger number of $100\%$ brighter than giant limit galaxies at lower redshifts, indicating a lack of faint galaxies even where they would be expected to be observable in low multiplicity systems. Error bars are calculated from the error in the mean. Right plot show the counts-corrected luminosity function for $N \leq 9$ and $N\geq 20$ groups. Blue depicts blue galaxies (as defined by the $b-r$ cut), red depicts red galaxies. Clearly the biggest discrepancy occurs at the faint end for early-types, the low multiplicity systems being massively depleted.}
\label{GroupComplete}
\end{figure*}

The suppression of the faint-end in the low-multiplicity systems can also be seen by plotting a variation on the dwarf-to-giant ratio (see Figure \ref{GroupComplete}), or directly from the luminosity distributions themselves (right hand panel). In the former, the fraction of galaxies brighter than the limiting absolute magnitude at the high redshift extreme of the survey volume ($M_{b_{j}}=-18.25$ at $z=0.1$) in each group is plotted as a function of group multiplicity and redshift. The lowest multiplicity systems maintain a larger fraction of galaxies with $M_{b_{j}}\leq -18.25$ to lower redshifts than their larger multiplicity counterparts, a large number still have fractional giant population of $\simeq 1$ (i.e.\ no dwarfs) at $z=0.07$. At such redshifts the limiting magnitude of the survey is comfortably fainter than this dividing line; assuming a typical galaxy k-correction it will be $\simeq -17.25$. Thus if the standard 2dFGRS Schechter LF values of \citet{coll01} were used, the expected population fraction of giants would be only 0.58 at this redshift. The point at which this fraction consistently reaches 0.5 is only at $z \sim 0.055$ (i.e.\ when the limiting magnitude is $-16.75$) for $N=5$ groups. In comparison $N=10$ groups achieve this level at $z \sim 0.07$, as expected. The overall luminosity distributions in the right hand plot demonstrate visually the clear differences in faint end slopes between small and large multiplicity groups for red galaxies in particular. 
 
\subsection{Field LF and Lower Multiplicities}

Recall that the low mass groups are a more specific environment than the general field (where local densities are computed over much larger volumes than
the typical group sizes). It is therefore not necessary (and indeed it is
not the case) that the faint end of the LF in such groups should tend to that of the field.  Field LFs tend to have $\alpha \sim -1.1$
rather than the shallower slopes for small groups seen here.

Despite the non-trivial nature of the `field', it is instructive to compare the trends discussed above with the LF parameters for all objects that are not in 2PIGG groups of multiplicity 5 and greater but meet the volume limited criteria of our sample. For the un-grouped sample (i.e.\ $N=1$) we find $M^{*}=-19.16 \pm 0.04$ and $\alpha=-0.83 \pm 0.02$ and for the groups too small to meet our sample ($2 \leq N \leq 4$) we find $M^{*}=-19.34 \pm 0.06$ and $\alpha=-0.71 \pm 0.04$. It should be highlighted that the fit for $N=1$ is of much better quality than for $2 \leq N \leq 4$. Comparing these numbers to the green points in Figure \ref{MaxLikeMult} indicates the trend in $\alpha$, and the more subtle trend in $M^{*}$ continues for lower mass groups that don't make our sample. The empirical LFs, along with the best fitting Schechter functions, can be found on the left panel of Figure \ref{MatchLF} for the data cuts described above. As a reference the composite LF for all group with $N\geq5$ is shown, for which we find $M^{*}=-19.79 \pm 0.03$ and $\alpha=-1.01 \pm 0.02$. Noticeable in this plot is the excess of bright galaxies for the $N\geq5$ population, despite the smaller integrated population. This can be interpreted as cruder evidence of the variation in $M^{*}$ with the scale of environment (un-grouped through to grouped), however we are only breaking the population into 3 large subsets.

For the red population, selected as above, we find for $N=1$ groups that $M^{*}=-18.95 \pm 0.03$ and $\alpha=-0.23 \pm 0.03$, and for $2 \leq N \leq 4$ groups $M^{*}=-19.18 \pm 0.03$ and $\alpha=-0.23 \pm 0.03$. Again, these values continue the strong trend in $\alpha$ into lower multiplicity/ mass groups. Finally, for the blue galaxies we find for $N=1$ groups $M^{*}=-19.49 \pm 0.03$ and $\alpha=-1.23 \pm 0.02$, and for $2 \leq N \leq 4$ groups $M^{*}=-19.62 \pm 0.04$ and $\alpha=-1.17 \pm 0.03$. Neither of these values varies significantly from that lowest mass groups plotted in figure \ref{MaxLikeMult}, and it is reasonable to conclude that any evolution of the blue galaxy luminosity function is weak or non-existant at low masses/ multiplicities.

\section{LF For Yang-2dF Groups}

The Yang-2dF group catalogue is based on the same survey data as the 2PIGG catalogue. However, as discussed in Section 2, the method for grouping the galaxies is markedly different. It is therefore instructive to see how different the best fit luminosity functions are for the two grouping methods.

In so far as is possible the same composite grouping criteria have been used: a requirement of at least 5 group members, the same survey volume and a division based on the total mass of the groups -- though this is now based on the method described by Yang et al. (2005b). Due to the poorer number statistics for the Yang-2dF groups the data has been split into 5 composite groups for each method of ordering, as opposed to the 10 used for 2PIGG.
 
\subsection{Similar Versus Non-Similar Groups}

As mentioned previously, a certain fraction of groups in the Yang-2dF and 2PIGG catalogues contain a large fraction of overlapping galaxy populations. The first obvious comparison to make, therefore, is between those groups that appear in both catalogues and that have the closest matching group centre. This latter requirement is so that we can ensure the comparison made is between the same central haloes, and means in the 47 cases where multiple Yang-2dF groups significantly match to a 2PIGG, only the closest groups in comoving space are compared. Obviously there is ambiguity in how best to match and compare such catalogues, but the aim here is to compare how the extra non-core galaxies affect the LF of the whole group- knowing that the Yang-2dF groups will be describing the core population only.

The best fit LF for matched groups based on the 2PIGG catalogue is $M^{*}=-20.07 \pm 0.05$ and $\alpha= -1.14 \pm 0.03$, compared to $M^{*}=-20.28 \pm 0.07$ and $\alpha=-1.16 \pm0.04$ for the matched Yang-2dF groups. The shapes of the LFs for the similar groups in the two catalogues agree best at the bright end and diverge slightly as a function of faintness, though evidently without a huge effect on the LF parameters as a large fraction of the galaxies in these two subsets will be the same. The biggest difference is simply in the normalisation, and suggests that where the two algorithms agree the LF is not varying greatly between the group core (reflected by Yang-2dF groups) and the extended infall region (captured by 2PIGG).

\begin{figure*}
  \centerline{
  \mbox{ \includegraphics[width=3.00in]{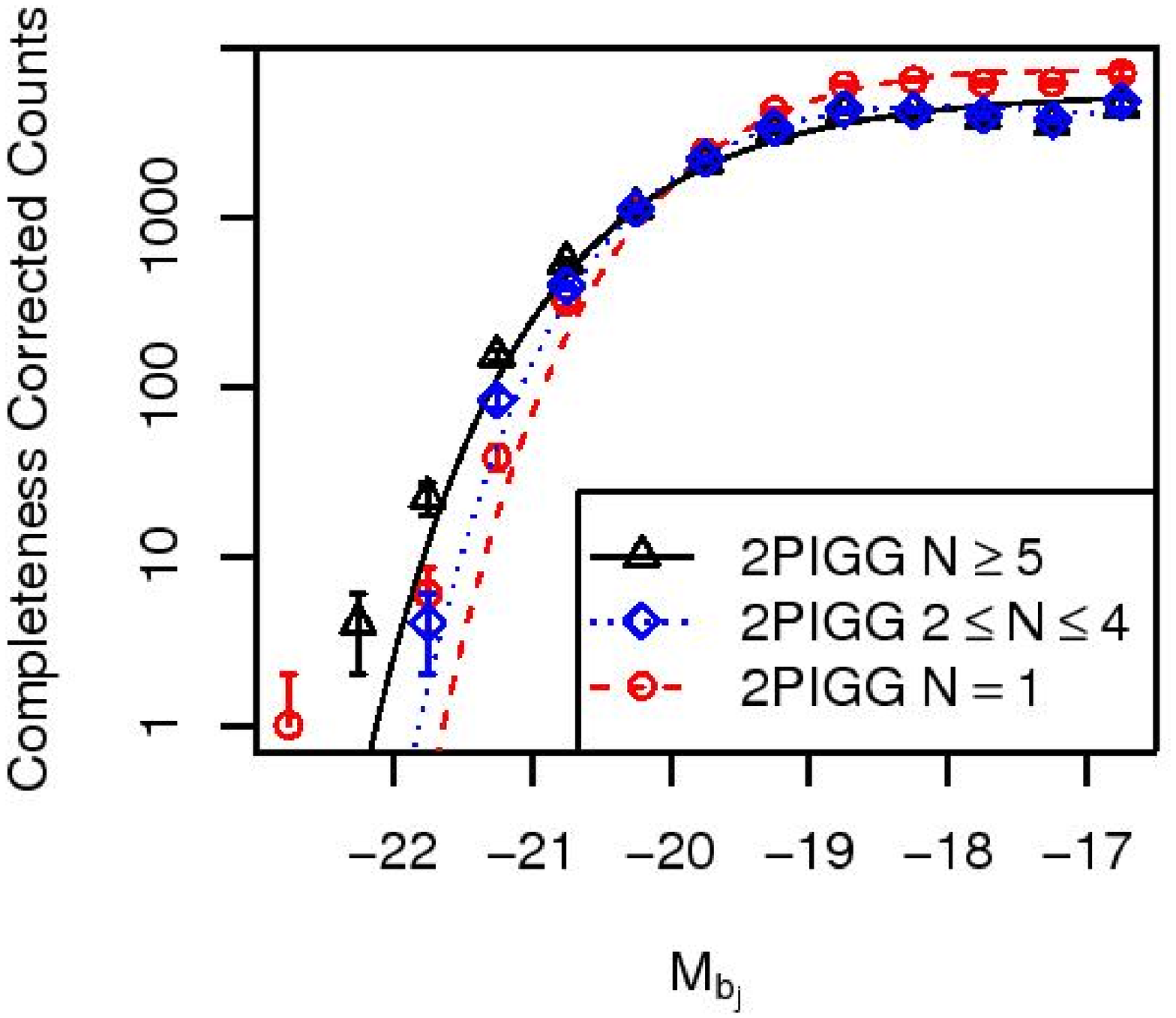}}
  \mbox{ \includegraphics[width=3.00in]{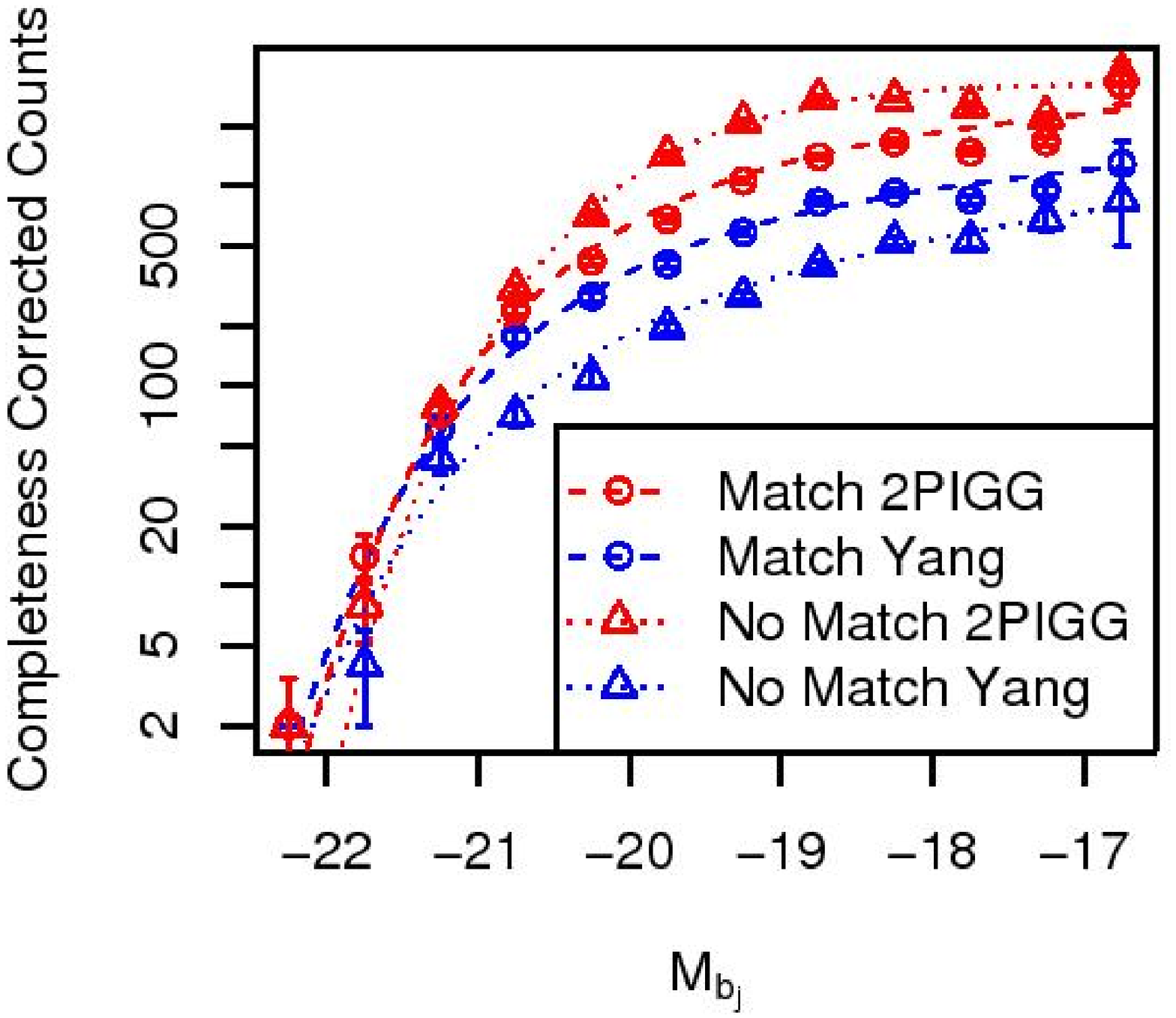}}
  }
  \caption{\small Left panel compares the LF for all 2PIGGs used in our sample, 2PIGGs that are grouped but are not in our sample by virtue of multiplicity ($2\leq N \leq 4$) and all un-grouped galaxies. Comparison of LFs formed by groups that are matched in the 2PIGG and Yang group catalogues, and groups which are unmatched. Data and best fit LFs are as depicted in the legend.}
  \label{MatchLF}
\end{figure*}

The remaining unmatched groups have a $M^{*}=-19.63 \pm 0.03$ and $\alpha=-0.92 \pm 0.02$ for 2PIGG, and $M^{*}=-20.42 \pm 0.12$ and $\alpha=-1.28 \pm 0.06$ for Yang-2dF groups. It is worth noting that the best fit LFs for these subsets generally have a low p-value ($\leq 0.1$), so it is dangerous to consider the best fit parameters alone. The best fit LFs along with the binned version of the data can be found in Figure \ref{MatchLF}.

A point of interest is that the matched and unmatched LFs for the same catalogue generally agree (or more specifically the LFs converge) at the bright and faint ends of the LF. In between these extremes the LFs diverge considerably. In the case of 2PIGG, the unmatched groups have a considerable excess of galaxies in the moderate luminosity ($-20\leq M_{b_{j}} \leq -19$) regime. In contrast, this is the region where the unmatched Yang-2dF groups display a relative deficit. The brighter $M^{*}$ value for the unmatched Yang groups is understandable since these groups (due to the method used to group them) will typically contain a dominant BCG. Since 2PIGG has no such bias, it is to be expected that in situations where the two catalogues fail to match a group the Yang groups will possess a brighter central galaxy.

 The unmatched group LFs are important because they give an insight into the most divergent behaviour between the grouping algorithms. The regions that have a relative excess of moderate luminosity galaxies will generally be  preferentially grouped by 2PIGG, and the reverse appears to be true for the Yang-2dF catalogue. This might indicate that the linking length used in 2PIGG for clustered moderate-luminosity galaxies is relatively generous. By this we mean that the global LF used for determining grouping over-estimates the rarity of moderate luminosity galaxies in dense environments,  these galaxies are apparently the key objects that determine successful grouping. In contrast, the lack of this moderate luminosity excess in the non-matched Yang-2dF groups implies that these galaxies do not add as much mass to the system (as calculated by the CLF), and do not increase the group-scale linking length as significantly. Ideally no one type of galaxy should dominate proceedings in this manner, but it does indicate that the groups with dominant central galaxies or very dense regions (i.e.\ regions the Yang et al.\ algorithm is designed to find) are relatively lacking in moderate luminosity galaxies, the expected signature of `cannibalism'. Obviously this indicates environmental dependence, but more generally this demonstrates that the Yang et al.\ algorithm is more prescriptive (compared to 2PIGG) about how the groups are built. This is reflected by the smaller variation in LF parameters between the similar and non-similar Yang-2dF/2PIGG systems.

\subsection{Mass Ordered Yang-2dF Groups}

\begin{figure*}
 \centerline{
    \mbox{\includegraphics[width=3.00in]{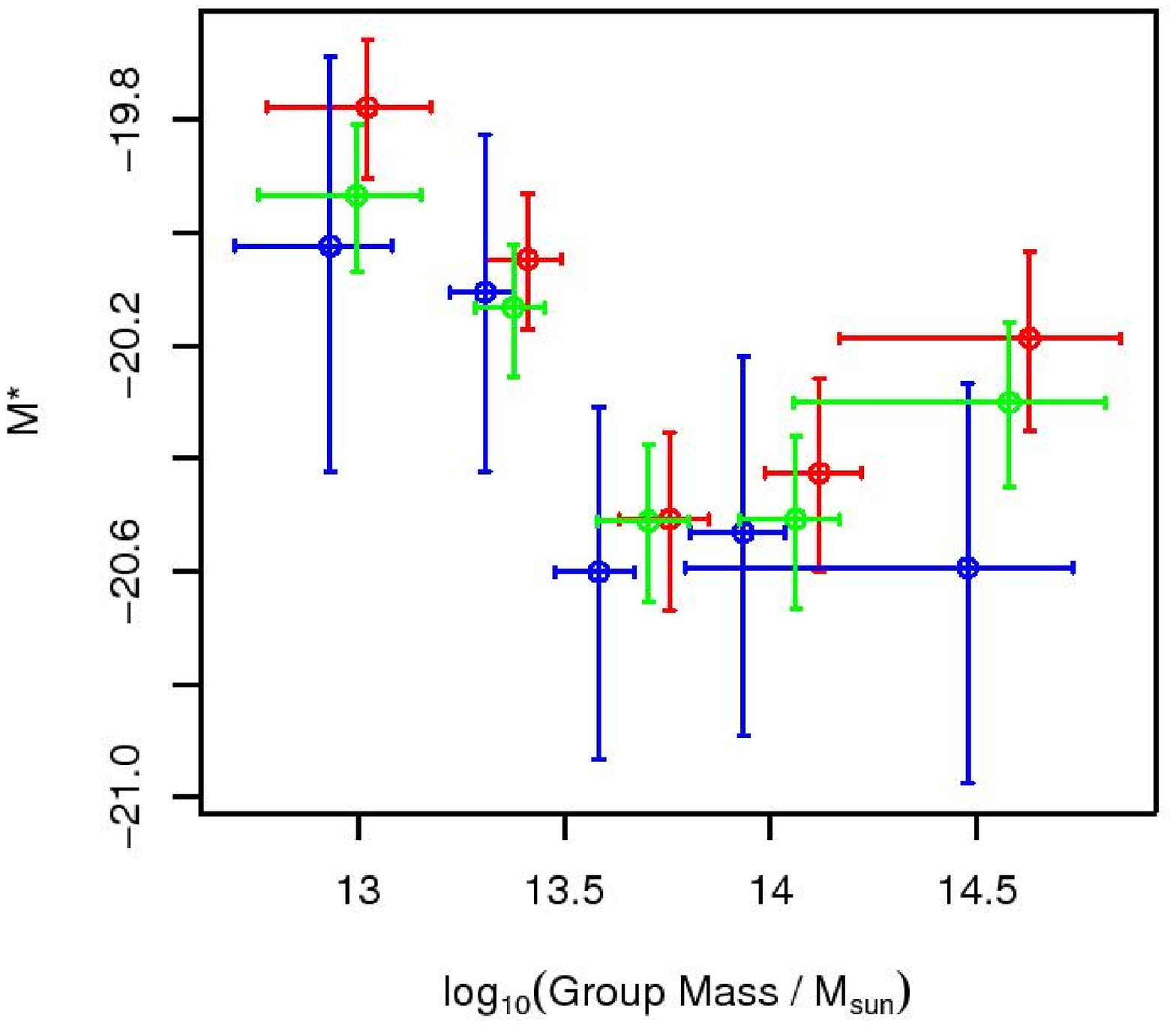}}
    \mbox{\includegraphics[width=3.00in]{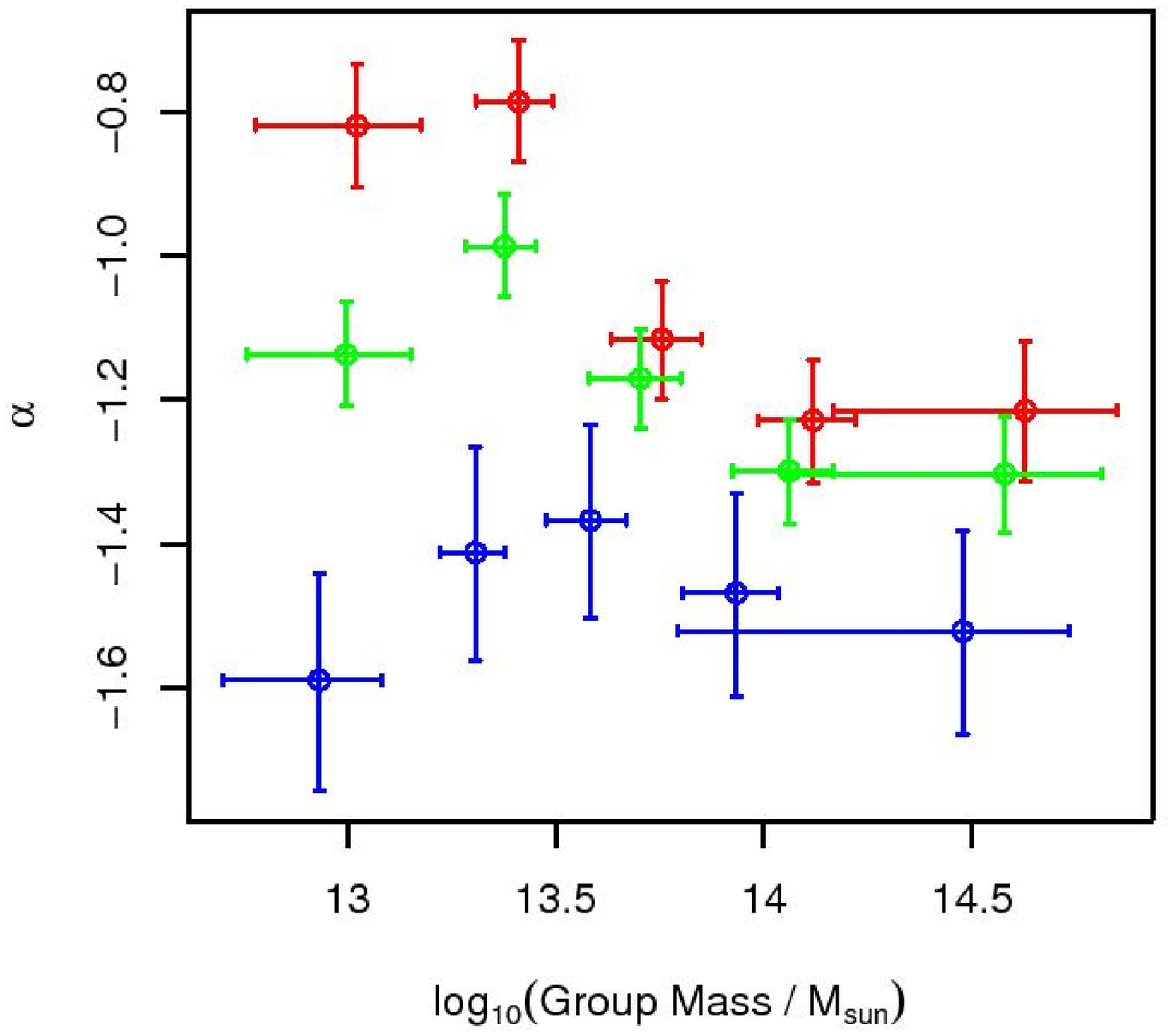}}
  }
 \caption{\small  Luminosity functions for Yang-2dF groups as found via Maximum Likelihood for blue, red and all galaxies (blue, red and green colours respectively). Plots show best fit $M^*$ and $\alpha$ for groups sorted into 5 bins via total mass estimates. x-axis errors are the standard deviation of the data, and the y-axis errors are the $95\%$ confidence levels computed from the log-likelihood.}
  \label{MaxLikeMassYang}
\end{figure*}

The luminosity functions ordered by total mass can be found in Figure \ref{MaxLikeMassYang}. The main variation compared to 2PIGG would seem to be steeper values for $\alpha$ and brighter values for $M^{*}$. The trends for $M^{*}$ appear to be much better defined than for the mass ordering of 2PIGG. However this is largely due to differences in how the total masses of the groups are calculated. The 2PIGG method assumes a degree of virialisation and is based on the velocity dispersion and a characteristic radius, while the Yang et al. method is based on the luminosity of the groups and an assumed mass-to-light ratio. As such it is largely the same as ranking by the total giant magnitude, as was done for 2PIGG in Robotham et al. (2006).

The combined, blue and red galaxies are all significantly brighter in $M^{*}$ for all masses compared to 2PIGG (though with quite large error bars). The red galaxies are found to have much steeper best fits throughout the range of composite groups compared to 2PIGG: the range covered is $-0.8 \geq \alpha \geq -1.2$ as opposed to $-0.3 \geq \alpha \geq -1$ for 2PIGG. The late types show no significant variations in $\alpha$, but are consistent with 2PIGG results given the large errors. The `all galaxies' LFs show a modest steepening at all masses, suggesting there is notable variation in how early-type galaxies are treated. 

\subsection{Multiplicity Ordered Yang-2dF Groups}
 
\begin{figure*}
  \centerline{
    \mbox{\includegraphics[width=3.00in]{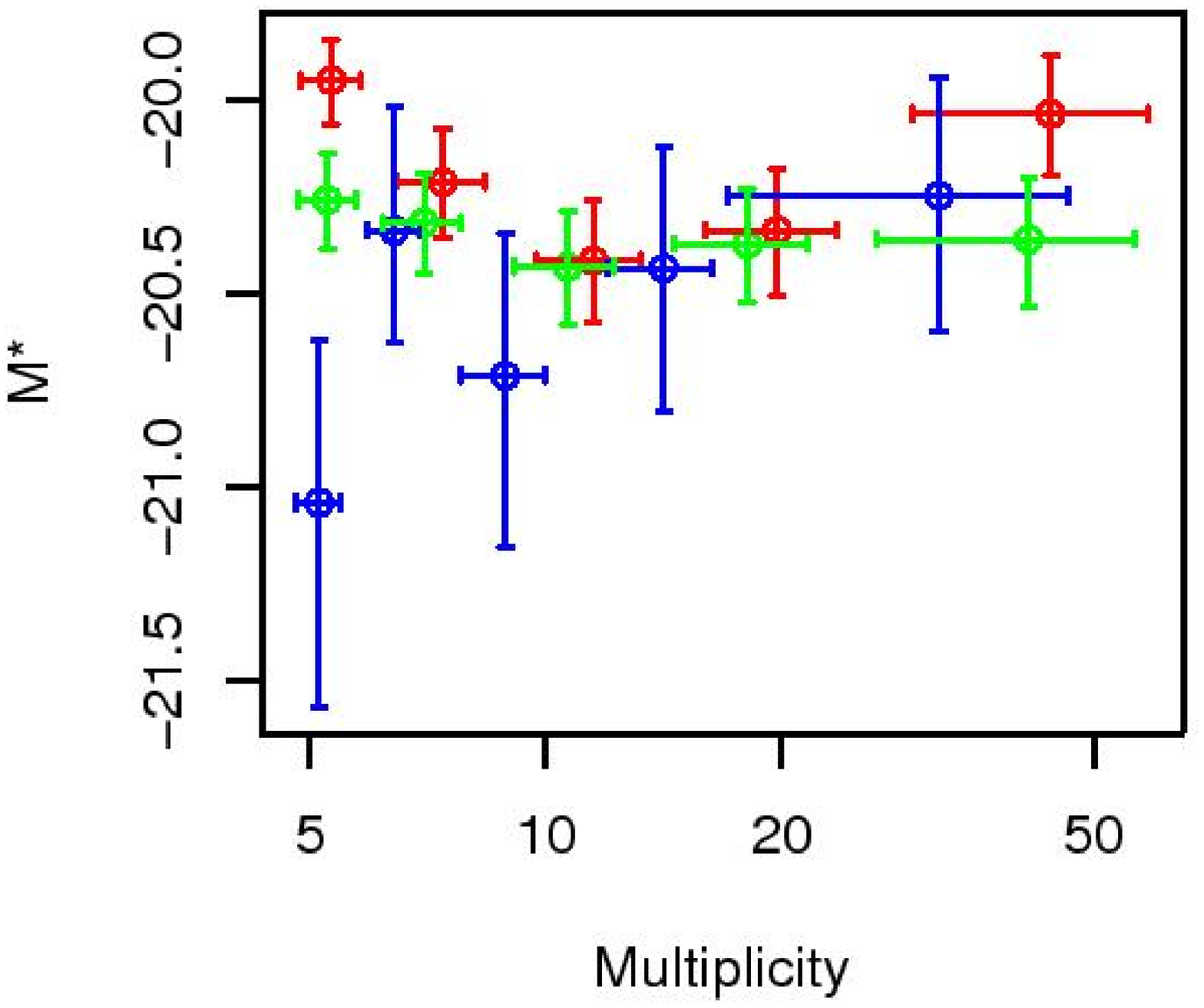}}
    \mbox{\includegraphics[width=3.00in]{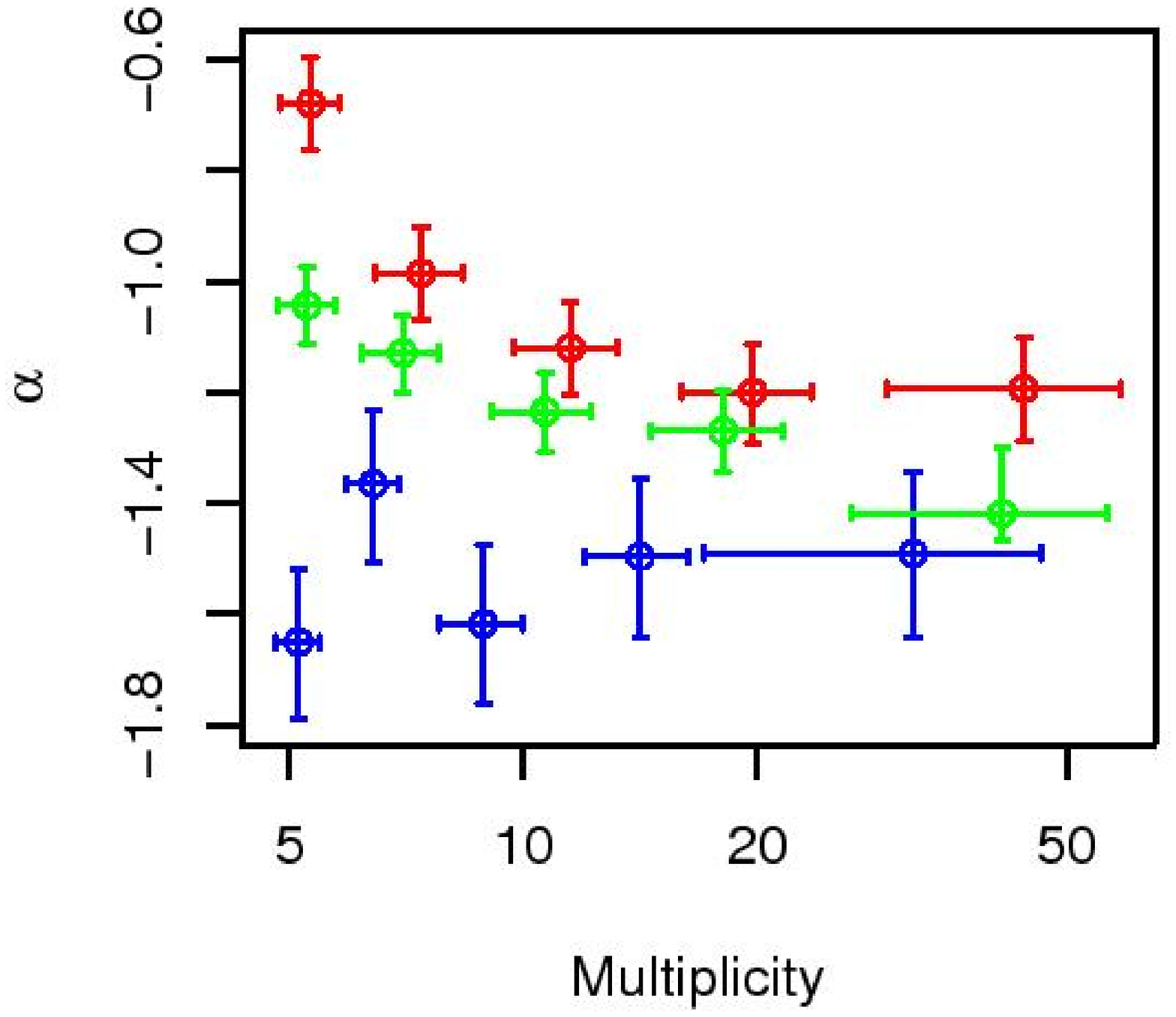}}
  }
  \centerline{
    \mbox{\includegraphics[width=3.00in]{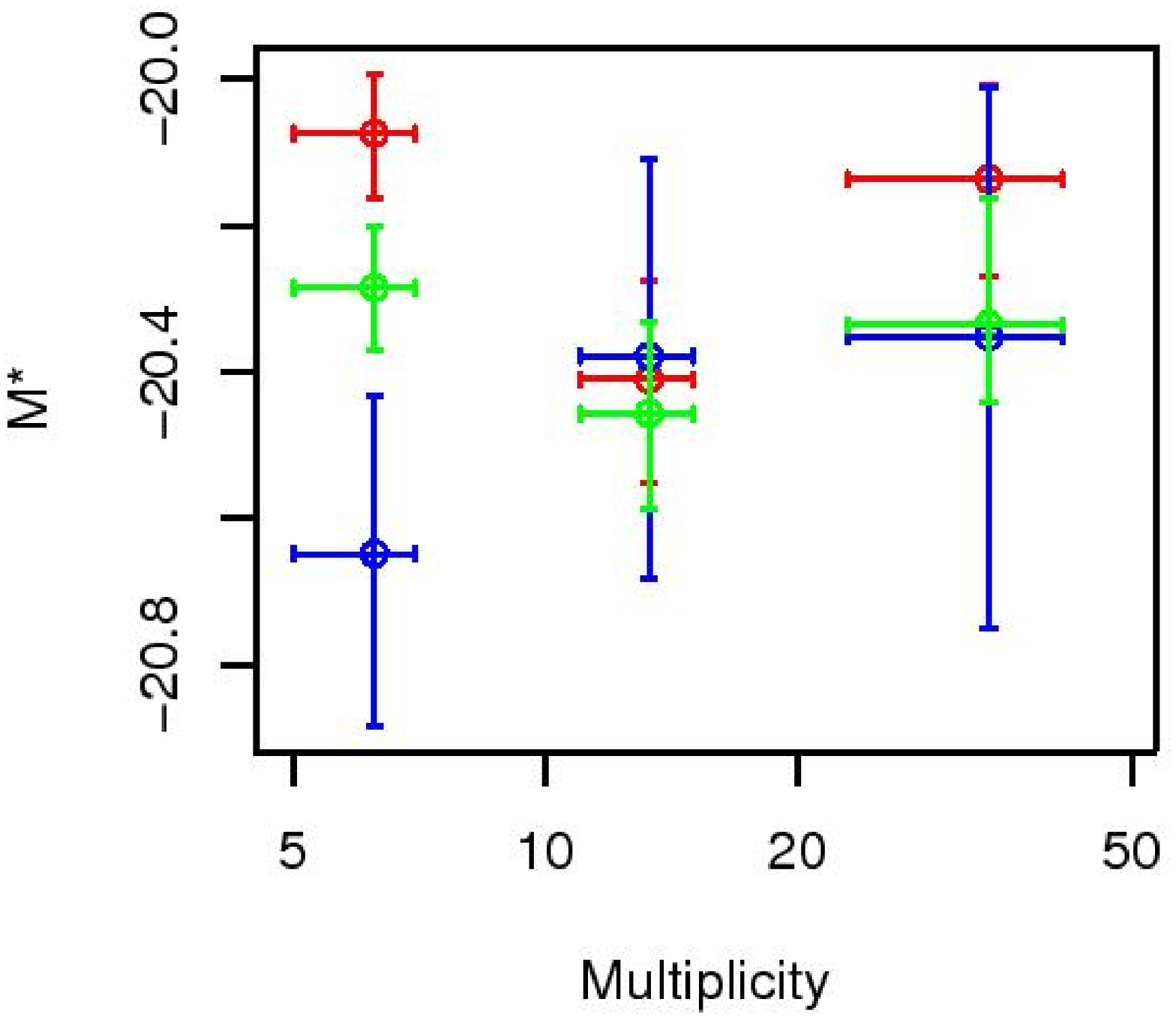}}
    \mbox{\includegraphics[width=3.00in]{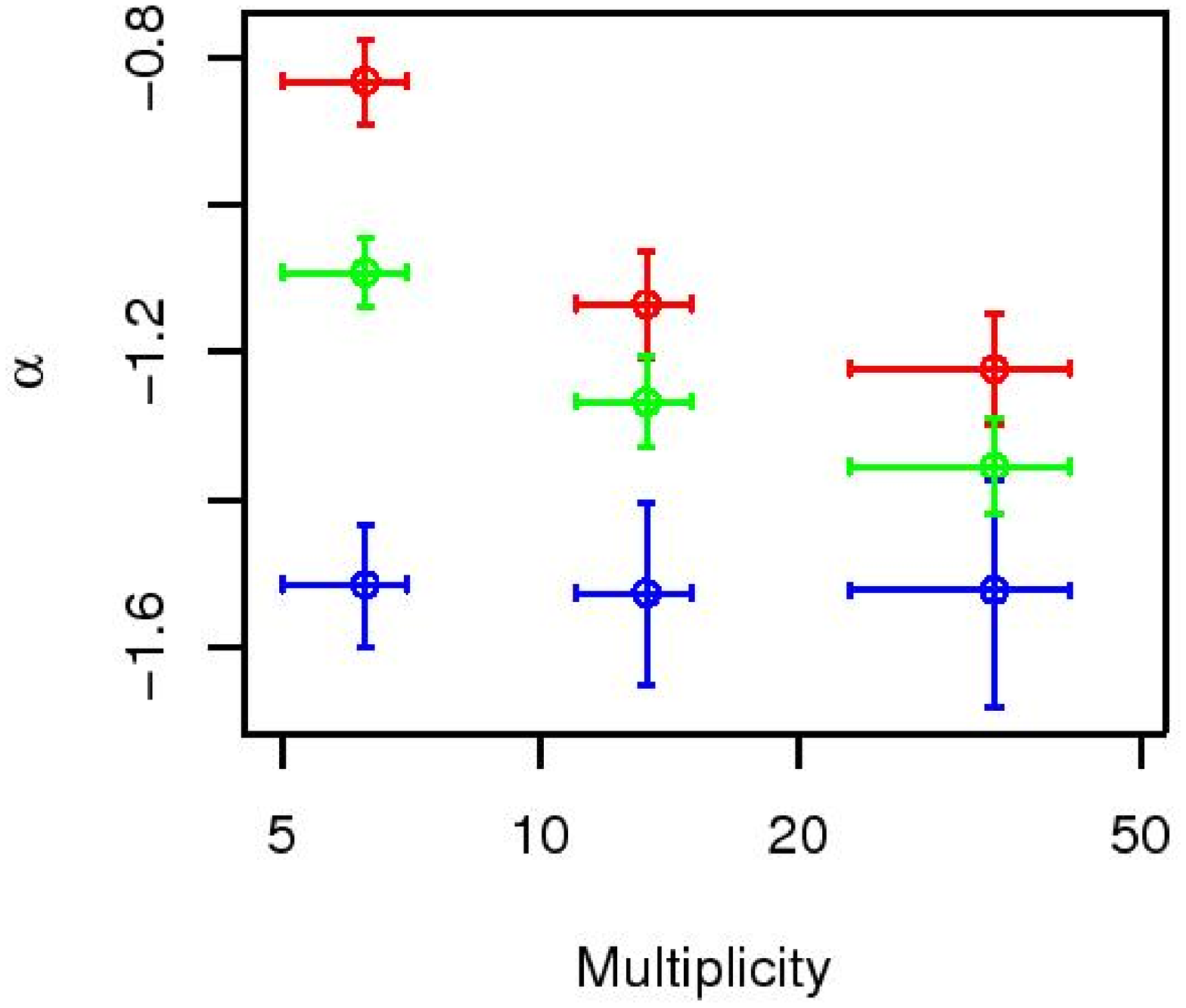}}
  }
  \caption{\small  Luminosity functions for Yang-2dF groups as found via Maximum Likelihood for blue, red and all galaxies (blue, red and green colours respectively). Plots show best fit $M^*$ and $\alpha$. Both top and bottom plots are sorted by group multiplicity. The top plot has 5 splits based on maintaining galaxy count in bins. The bottom plot has three bins of roughly similar galaxy count, but specifically the 3 ranges: $5\leq N\leq 9$, $10\leq N\leq 19$ and $N\geq20$. x-axis errors are the $25\%$ and $75\%$ quantiles, and the y-axis errors are the $95\%$ confidence levels computed from the log-likelihood.}
  \label{MaxLikeMultYang}
\end{figure*}

As for 2PIGG, the other method used for ordering the data was by group multiplicity. The results for this can be found in Figure \ref{MaxLikeMultYang}, the top plots show the best fits for 5 composite groups with efforts to maintain galaxy numbers in each composite group, the bottom plots use the same fixed multiplicity limits as previously used for 2PIGG ($5\leq N\leq 9$, $10\leq N\leq 19$ and $N\geq20$).

For the 5 bin split via multiplicity, the errors for the late-types are typically very large, so the apparent faintening trend as a function of multiplicity may not be reliable. The change in $M^*$ for early-types for the lowest multiplicities again appears to be present in this data, though as with all the Yang-2dF composites, it takes a brighter value than for 2PIGG.

The best fit values for $\alpha$ are again generally steeper than for 2PIGG for equivalent composite group subsets, but the trends are much the same. The early-type galaxies show very clear steepening as a function of multiplicity and, much like was found for 2PIGG, the low multiplicity systems show the most dramatically depressed faint end with $\alpha \sim -0.7$. The blue galaxies have $\alpha \sim -1.5$ throughout.

\subsection{LF Differences Between Grouping Algorithms}

Whilst there is a good deal of agreement between 2PIGG and Yang-2dF group luminosity functions, it is the differences that are probably the more interesting. As discussed above, the most distinct variation appears to be that $\alpha$ values are routinely steeper for the Yang-2dF groups than for 2PIGG. This is worth investigating in more detail.

\begin{figure*}
 \centerline{
   \mbox{\includegraphics[width=3.00in]{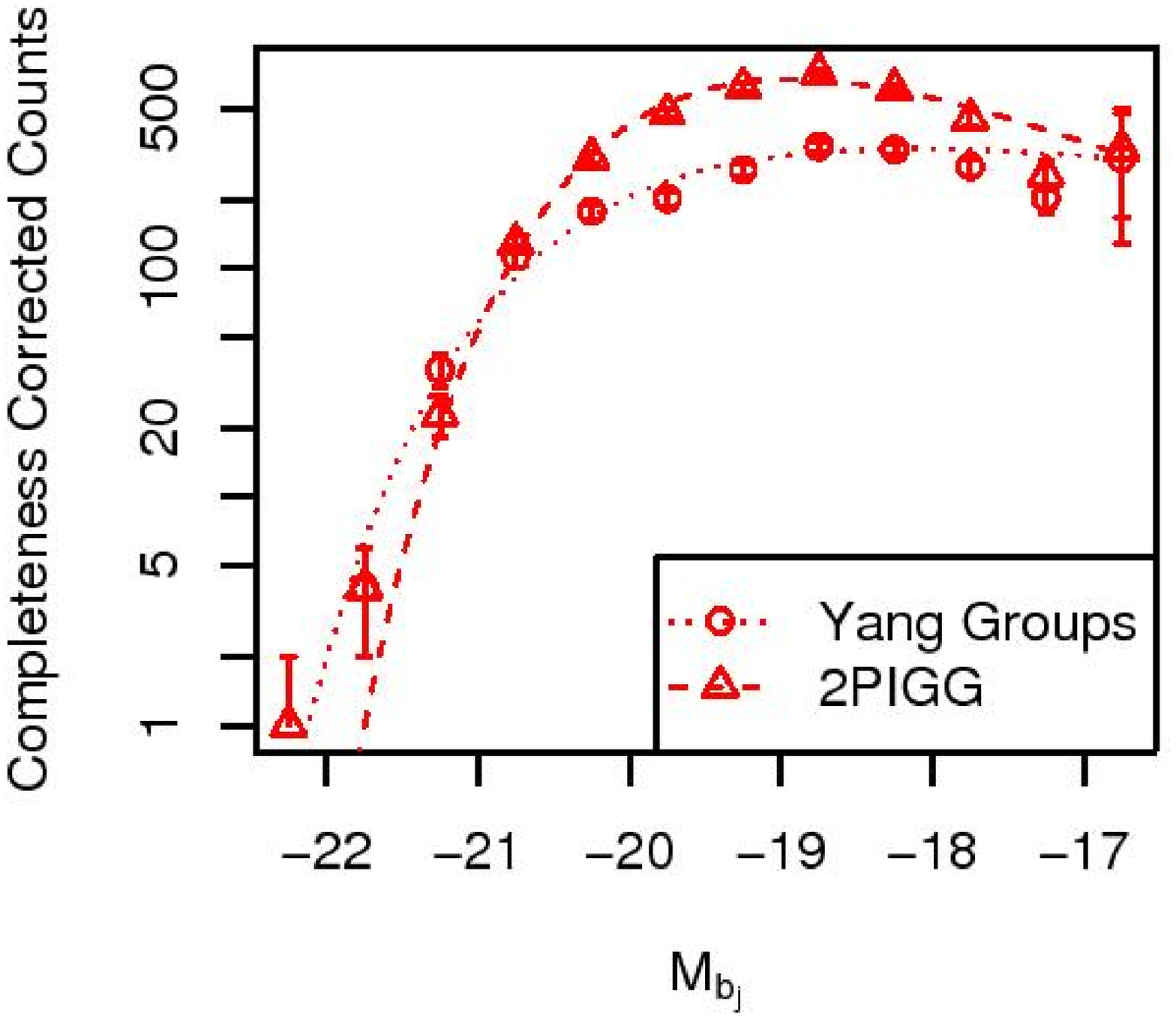}}
   \mbox{\includegraphics[width=3.00in]{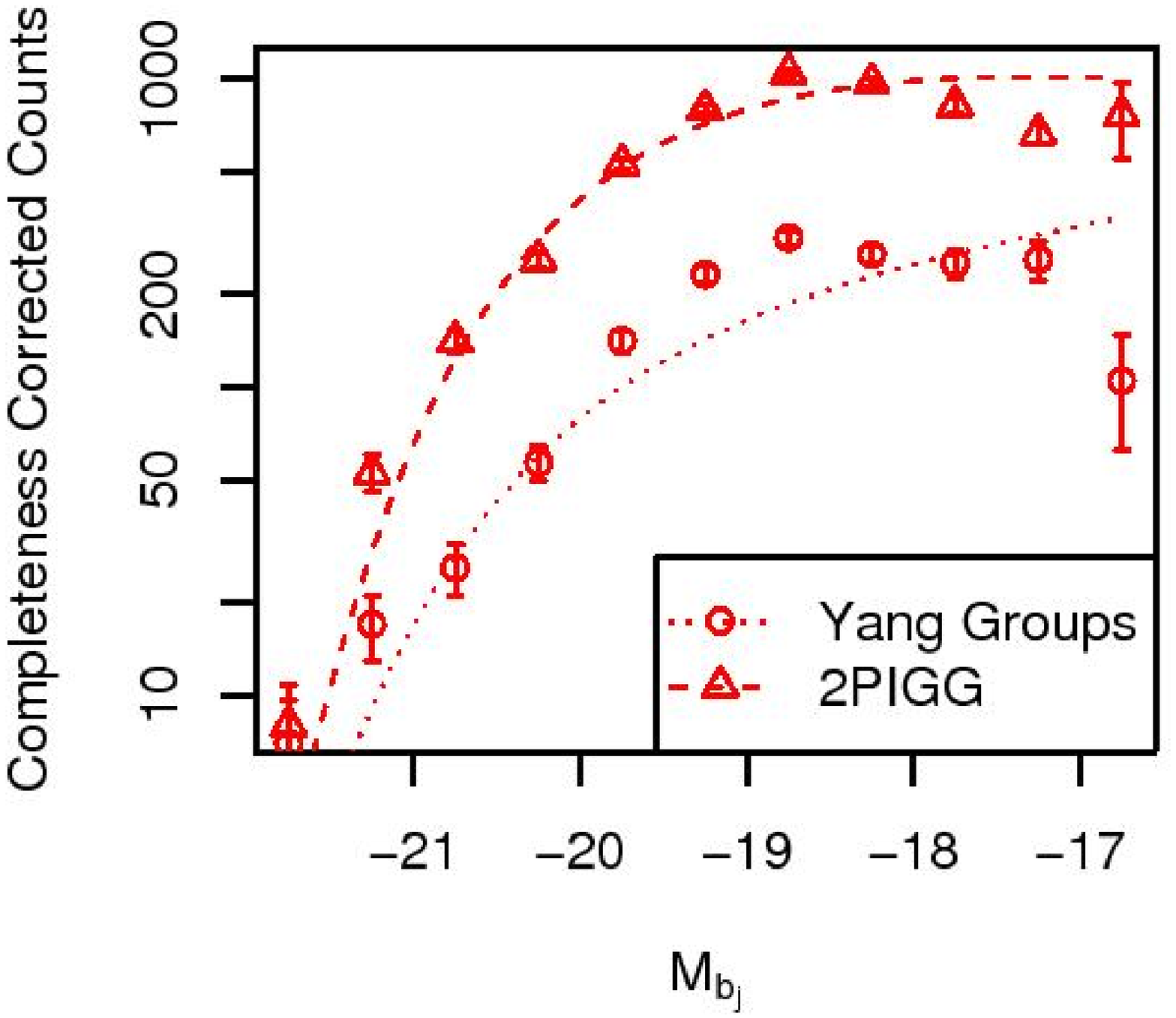}}
}
\caption{\small  Red galaxy LFs for Yang-2dF groups and 2PIGG. Shown are the completeness corrected counts and the best fit maximum likelihood LF. The low multiplicity ($5\leq N \leq 9$) comparison are shown in the left plots, the large multiplicity ($20\geq N$) comparisons on the right. Disagreement between the data and best fit is partially caused by the use of maximum likelihood (not binned data) and because the mid-point of the bin is used for scaling purposes.}
\label{EarlyLFComparison}
\end{figure*}

\begin{table}
\begin{center}
\begin{tabular}{l||rr|rr}
\multicolumn{1}{c||}{} &
\multicolumn{2}{c|}{2PIGG} &
\multicolumn{2}{c}{Yang-2dF}\\
\multicolumn{1}{c||}{Multiplicity} &
\multicolumn{1}{c}{Red} &
\multicolumn{1}{c|}{Blue} &
\multicolumn{1}{c}{Red} &
\multicolumn{1}{c}{Blue}\\ \hline
5-9 &3057&3459&1610&838\\
10-19 &2374&2019&1019&449\\
20+ &4248&2648&1174&279\\
\end{tabular}
\end{center}
\caption{\small  Table comparing the numbers of different galaxy types (red/blue) for 2PIGG and Yang-2dF groups for different multiplicity subsets.}
\label{MultCompTab}
\end{table}
Figure \ref{EarlyLFComparison} is a like for like comparison between the two catalogues at the small ($5\leq N \leq 9$) and large ($20 \leq N$) extreme for red galaxies, and it is immediately clear that whilst the two populations display differences for both of these multiplicity subsets, it is in the lowest multiplicity cut that the most significant discrepancies appear.

Considering the large multiplicities first, the Yang-2dF group data is more similar to 2PIGG in its LF than the best fit implies, as the underlying LF is not well reproduced by the fitted Schechter function (as reflected by a very low p-value for the best fit of 0.0014). The major disagreement between the two LFs is at the bright end, where the number of giants in the Yang groups is comparitively high and thus the dwarf-to-giant ratio is lower. Elsewhere the results indicate a similar PDF for the LF but a less rich population -- for most bins there is a factor $\sim 3$ difference. From the variation in gradient seen, it appears that the Yang-2dF groups would be better fit by a combination of separate giant and dwarf (satellite) populations (c.f. Popesso et al.\ 2005, for clusters). Specifically, the gradient steepens at $M_{b_{j}}\sim -20.5$, when for a pure Schechter function it will always decrease.

The $5\leq N \leq 9$ LFs show a great deal of disagreement between $-20 \leq M_{b_{j}} \leq -18$, where 2PIGG displays a general excess of moderate luminosity galaxies compared to Yang-2dF groups. Interestingly, the two populations converge almost completely at the giant and faint extremes, suggesting that for these low multiplicity systems the most bound structure is at either extreme of the luminosity function, and the tenuous structure (as grouped by 2PIGG) tends to be reflected by the moderate luminosity red galaxies. Physically, 2PIGG may be `attaching' extra moderate luminosity objects which are/were the central galaxy of their own halo when grouping objects, whereas the Yang et al. algorithm does not. 

\begin{figure*}
 \centerline{
   \mbox{\includegraphics[width=3.00in]{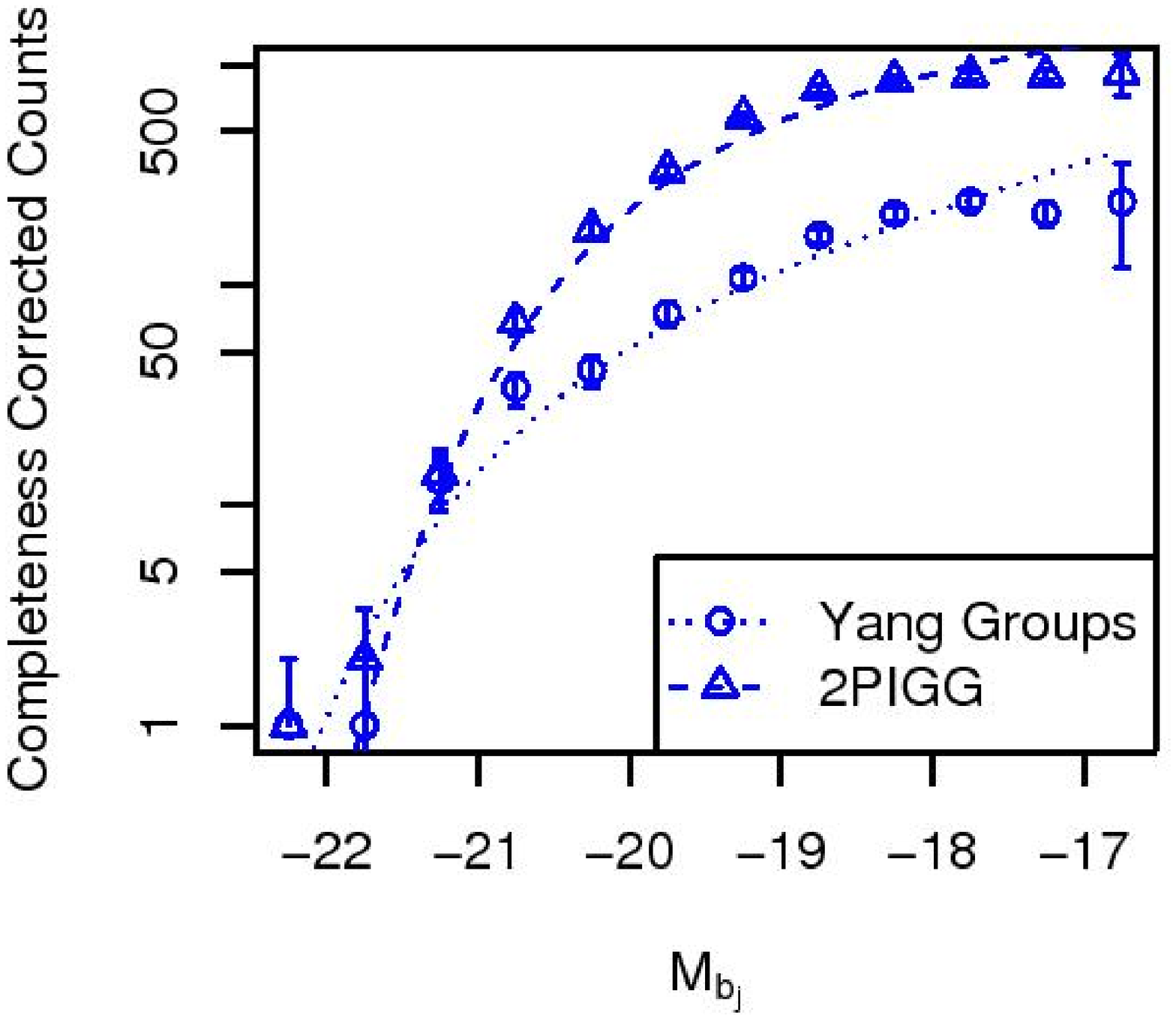}}
   \mbox{\includegraphics[width=3.00in]{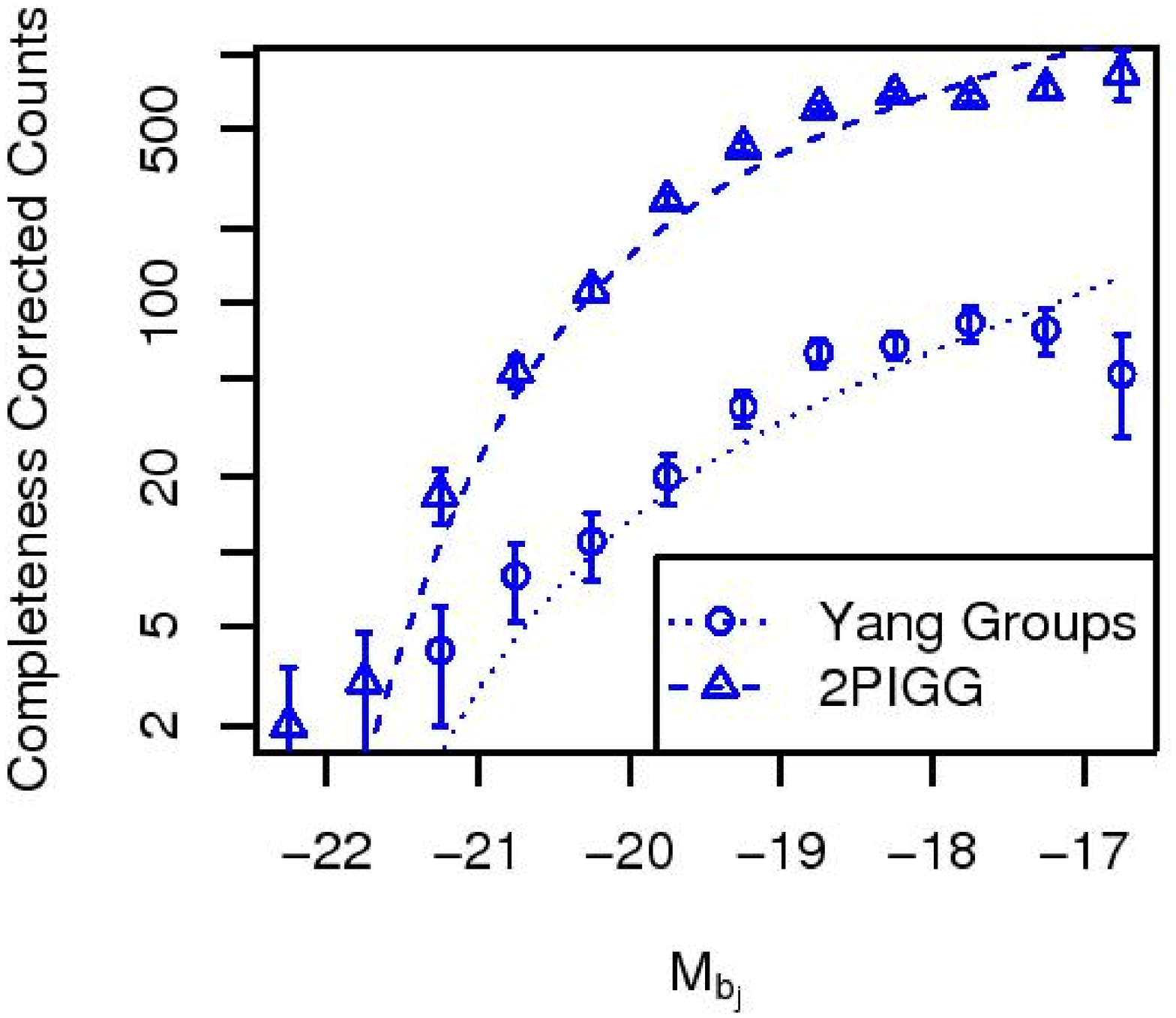}}
}
\caption{\small  Blue galaxy LFs for Yang-2dF groups and 2PIGG. Shown are the completeness corrected counts and the best fit maximum likelihood LF. The low multiplicity ($5\leq N \leq 9$) comparison are shown in the left plots, the large multiplicity ($20\geq N$) comparisons on the right. Disagreement between the data and best fit is partially caused by the use of maximum likelihood (not binned data) and because the mid-point of the bin is used for scaling purposes.}
\label{LateLFComparison}
\end{figure*}

Figure \ref{LateLFComparison} displays the different LFs for blue galaxies using the same multiplicity subsets. The blue galaxies in Yang-2dF groups suffer poor number statistics for both multiplicity cuts, as implied by the larger error bars, and subsequently have a noisier distribution around the best fit. From Table 1, the lower multiplicity Yang-2dF groups are substantially less rich in blue galaxies than their 2PIGG counterparts (34\% of the galaxies compared to 53\%) and the high multiplicity systems are relatively even poorer in blue galaxies (19\% compared to 38\%).
Thus the ratio of red to blue galaxies is always larger for Yang-2dF groups, but this ratio increases dramatically as a function of multiplicity, indicating that in more bound systems (in this case, the Yang-2dF groups) a greater fraction of blue late-types (dominating the blue sample) have made the transition to early-types (dominating the red sample), even at fixed multiplicity. As these more bound groups will have higher density, this then reflects the usual \citet{dres80} morphology-density relation, but within fairly small groups. 

In addition, the blue galaxy LF remains fairly consistent in its PDF (i.e.\ shape), between the catalogues, whilst the red galaxy LF is altered as a function of both richness and the bounding criterion used. The halo-based method for the Yang-2dF groups clearly produces a relative deficit in moderate luminosity red galaxies compared to the FoF 2PIGG, particularly for low multiplicity systems. This may indicate that early-type galaxies possess a LF shape that varies significantly as a function of environment, i.e.\ environment affects different luminosity galaxies quite differently, whilst late-type LFs mostly vary in richness as a function of environment, the shape remaining relatively fixed.

There is a caveat to making strong conclusions regarding the discrepancy between 2PIGG and Yang-2dF groups early/late-type ratios, since it can be partially explained by a relative excess of interlopers in 2PIGG systems. These interlopers are likely to be field galaxies, and will tend to be late-type (blue) galaxies. Whilst this might partially explain the discrepancies in relative richness, it does not naturally explain variations in LF shape seen for the red galaxy sample.

\section{Position Dependence}

To explore this a little more we consider the variation of the luminosity function with position within `halos' for the 2PIGG catalogue. Taking the centres determined previously, we have spatially `stacked' the groups in the three broader multiplicity composites (to maximise the galaxy numbers) and obtained LFs for the combined and the separated blue and red populations, as defined above.

The LF was calculated for both types within a projected radius of 0.5 $r_{rms}$, and outside of 1 $r_{rms}$ for each multiplicity. These limits were chosen so as to make sure a large fraction of galaxies in these areas are physically within the 3D radii regimes of interest. The results for these fits can be found in table \ref{ProfLumTab}.
 
We can see that the outer regions maintain a similar LF shape through all environments and galaxy types, within 95\% confidence errors. The central LF for late-types does appear to change slightly as a function of multiplicity, however this variation is only just outside the confidence range for the parameters. The most dramatic variation is certainly that seen for early-type (red) centrally located galaxies. The key parameter of interest, since it will determine the majority of galaxy counts in a given volume, is $\alpha$ for the early-type galaxies. For the multiplicity 5--9 groups it changes from $\alpha=-0.52$ centrally to $\alpha=-0.21$ in the outer parts (the latter remaining similar for the other multiplicities), but the central LFs have $\alpha=-0.85$ for groups with 10--19 members and  $\alpha=-1.00$ for the largest systems. In the latter case the LF is actually quite a poor fit (p-value of 0.01), and a visual inspection of the raw LF indicates a two component model is almost certainly more appropriate, but regardless of this, the faint end is clearly markedly steeper for the 10--19 and 20+ member groups compared to 5--9 multiplicities. 

Following \citet{crot05}, this implies that the outer regions of groups are very similar environments and apparently influenced almost uniquely by the local (instantaneous) density of the environment, which will be similar at the outer limits of all FOF grouped system \citep[see also the discussion in][]{bald06}. This has the added implication that beyond $r_{rms}$, interactions with the denser cores of groups have had little impact on the galaxy population, else we would expect to see some variation in systems with much denser central regions. These results are consistent with the work of, for example, Haines et al. (2006), who found that the dwarf population demonstrated a sharp transition between early and late-type dominance near 1 virial radius from the centre of Abell 2199. The conclusion of that work was that dwarf galaxies only become passive once they have become satellites within a more massive halo. More generally, Popesso et al. (2006) suggest that galaxy transformation occurs at around $0.7 r_{200}$, and it would appear that similar conclusions can be drawn on scales much smaller than the clusters used in their work. Such a picture is consistent with semi-analytic models of star formation suppression \citep[see e.g.][]{bowe06,catt07}.

\begin{figure*}
\centerline{
  \mbox{\includegraphics[width=3in]{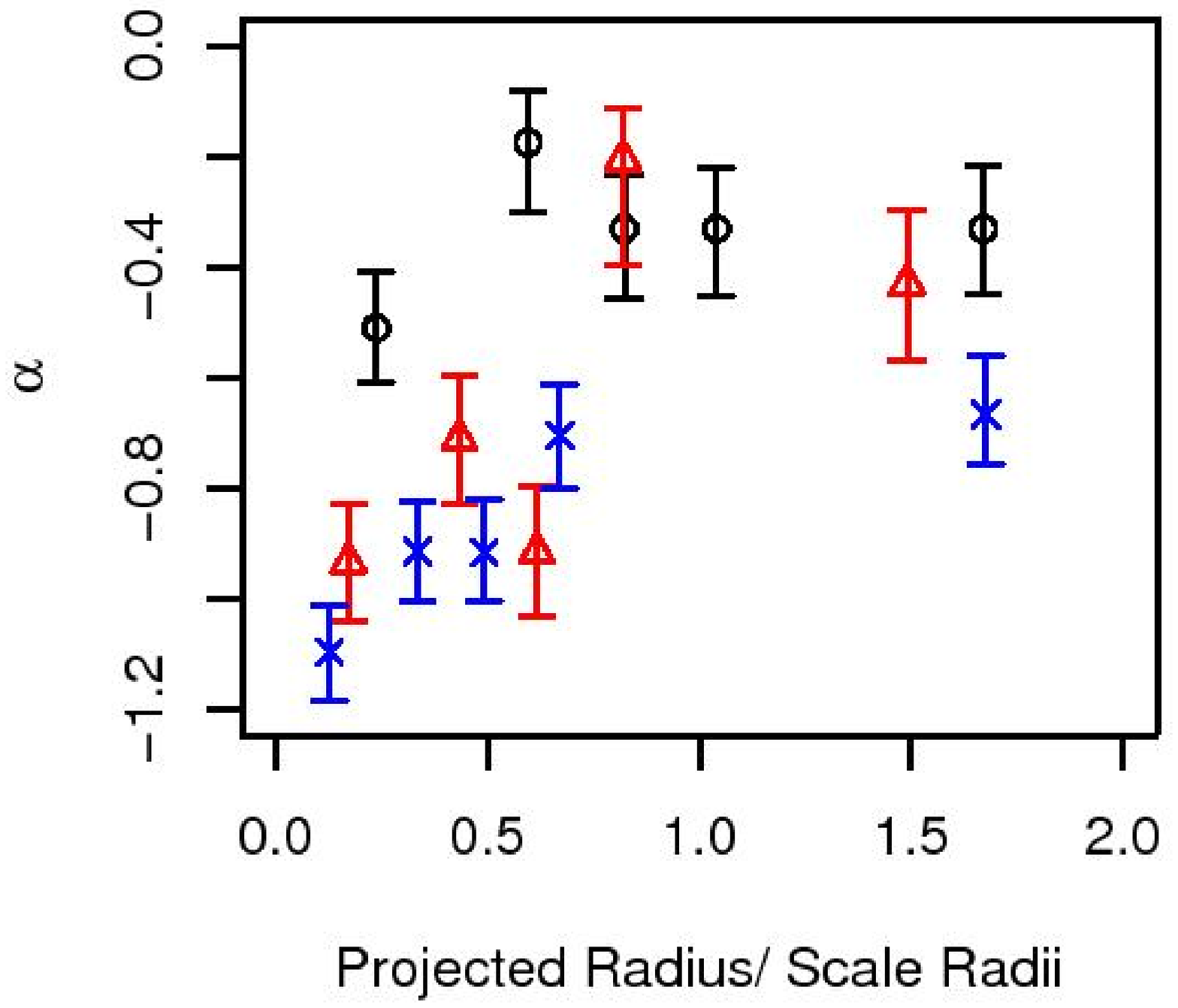}}
  \mbox{\includegraphics[width=3in]{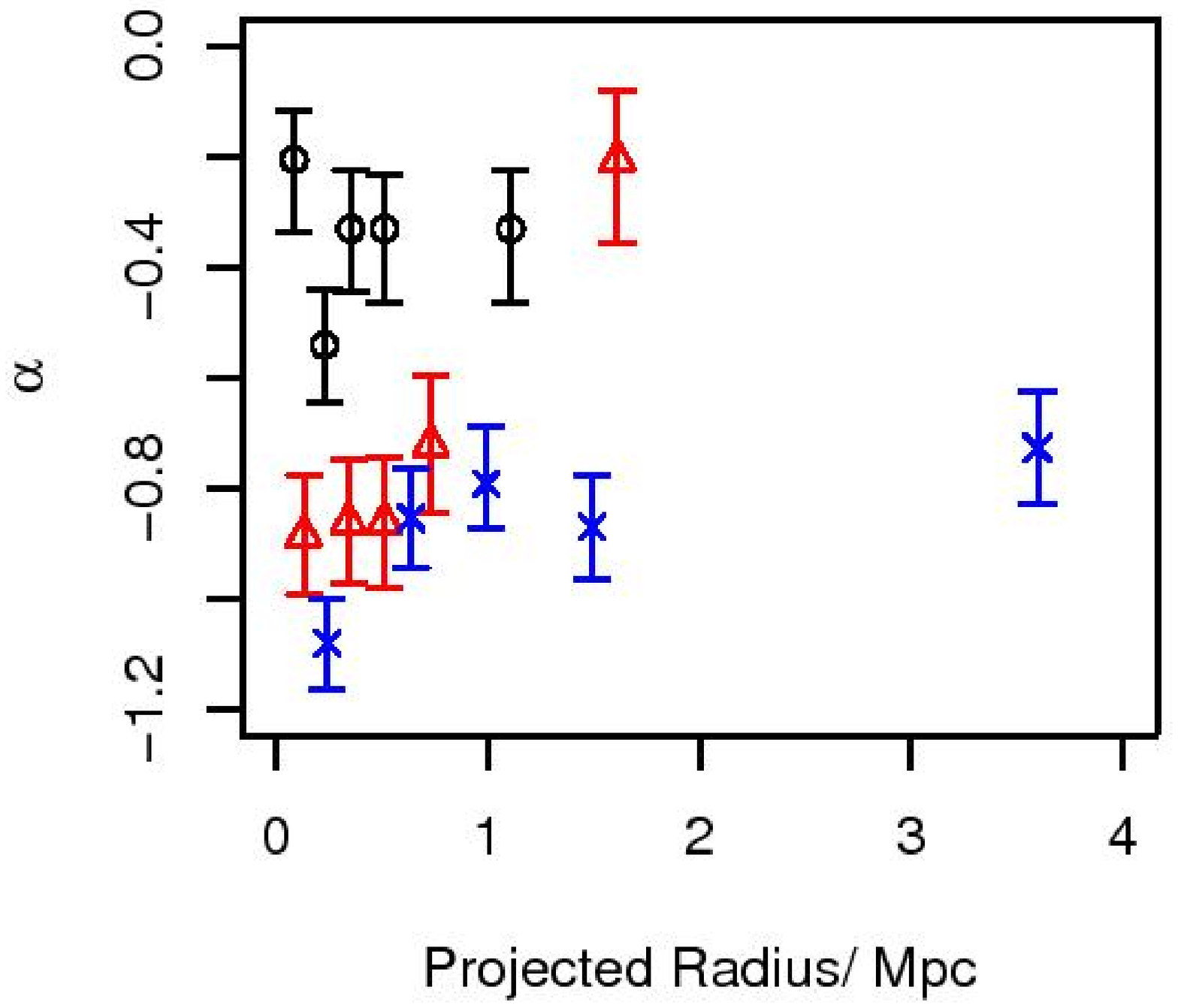}}
  }
\caption{\small The left-hand plot shows how $\alpha$ varies as a function of scaled radius $r/R_{rms}$ for multiplicity 5-9 (black circles), 10-19 (red triangles) and 20+ (blue crosses) groups. The right-hand plot keeps the same environmental divisions, but now as a function of the true projected radius $r$. In all cases the divisions are based on maintaining galaxy counts in the subset (hence similar 95\ confidence regions), and the midpoints of the bin is used for plotting purposes.}
\label{radialLF}
\end{figure*}

It is important to note that $r_{rms}$ appears to be the important threshold for defining the region where the LF dramatically changes shape, most significantly for early-type galaxies. Outside of this, grouped objects appear to share a largely similar LF, regardless of the size of the system (as determined by multiplicity). The strength of this effect is readily observable in Figure \ref{radialLF}, which compares $\alpha$ as a function of the scaled radius and the actual projected radius. It seems clear that the scaled radius is a better tool for finding regions with similar LFs, and all three environments used in this work are most similar beyond 1.0 $r_{rms}$. 

\subsection{Luminosity and Distance Ranks}

Related to these positional variations is the question of luminosity-distance correlations. If a group is well virialised and evolved, then the process of merging will tend to place the most luminous galaxies closest to the central galaxy. The definition of the group centre varies between 2PIGG and Yang-2dF groups. 2PIGG uses an iterative process to continually remove the galaxy furthest from the completeness weighted centre until only two are left. The final step is to select either the galaxy with the largest weight, or if identical the galaxy with the most $b_{j}$ flux. The Yang algorithm uses a luminosity weighted group centre from the outset, so at least the brightest member will have an enhanced tendency to be near to the defined group centre. We are interested in uncovering the luminosity gradient from the brightest group member rather than relative to a somewhat arbitrarily defined group centre. With this in mind all, group centres were redefined from the brightest galaxy for both 2PIGG and Yang groups, and biases caused by differing definitions of the group centre were therefore removed.

As a simple measurement, a real merger gradient implies a negative correlation between galaxy luminosity and distance from the brightest galaxy in the group (or a positive correlation if we use magnitudes). To remove the issue of group scaling, the Pearson product-moment coefficient was calculated for each group in turn (from both the 2PIGG and Yang-2dF catalogues), and this was compared to the distribution expected for randomly generated data with the same multiplicities. The form of the Pearson statistic ($P_{r}$) measured was

\begin{equation}
P_{r} = \frac {1}{n - 1} \sum^n _{i=1} \left( \frac{M_{r_i} - \bar{M_{r}}}{\sigma_{M_r}} \right) \left( \frac{r_i - \bar{r}}{\sigma_r} \right)
\end{equation}

\noindent where $M_{r_{F}}$ is the absolute $r_{F}$ band magnitude and $r$ is the projected distance from the brightest galaxy for each galaxy. This statistic returns 0 if the correlation is non-existant, 1 if it is perfectly linear and positive and $-1$ if it is perfectly linear and negative. All distances are relative to the brightest group member so the brightest galaxy is removed from the calculation, the remaining gradient is for the second rank galaxy and fainter. Since it is the projected distance that is used the correlation, if present, will be reduced in significance by the random positions of group foreground and background galaxies.

\begin{figure*}
\centerline{
 \includegraphics[width=3in]{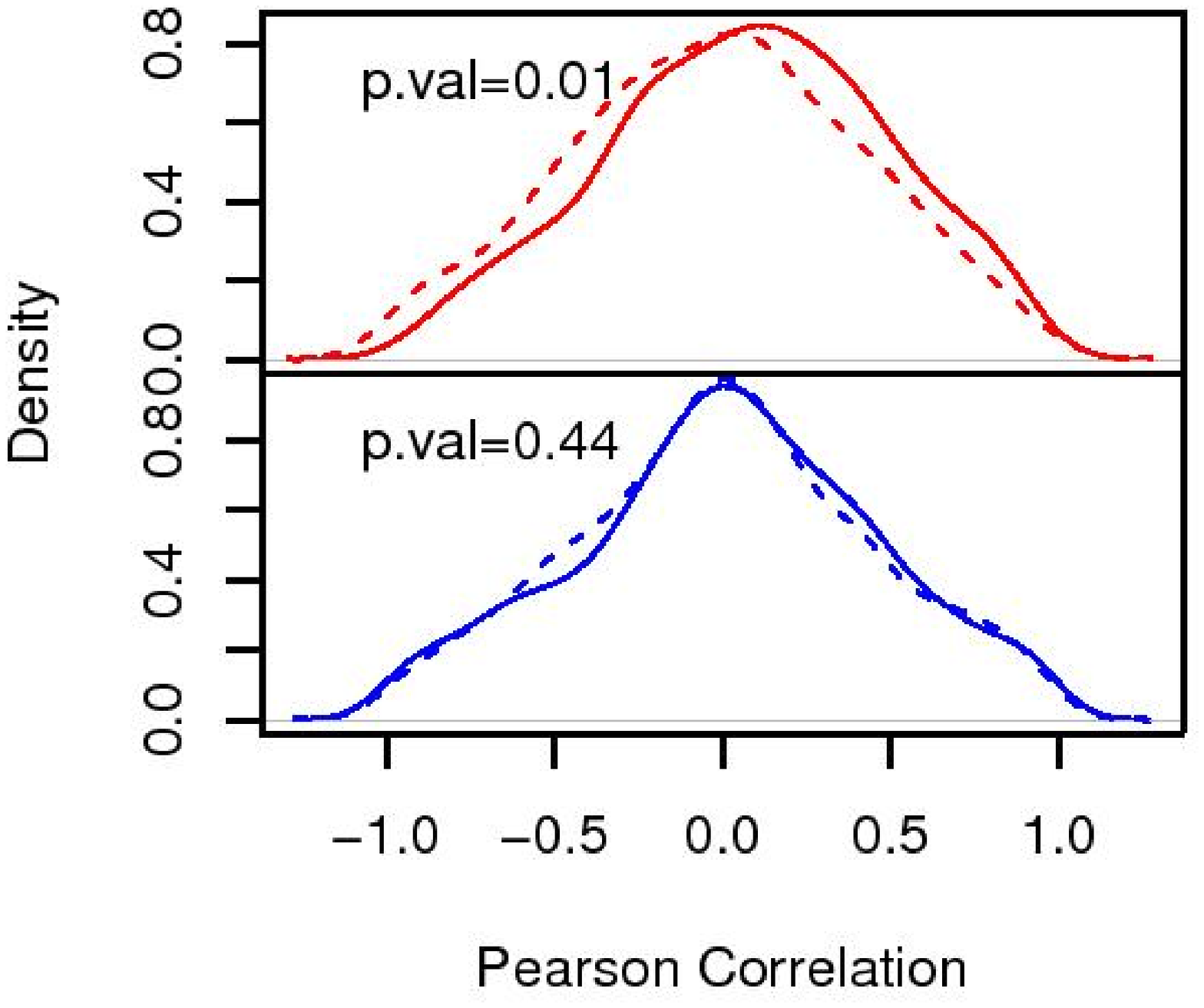}
 \includegraphics[width=3in]{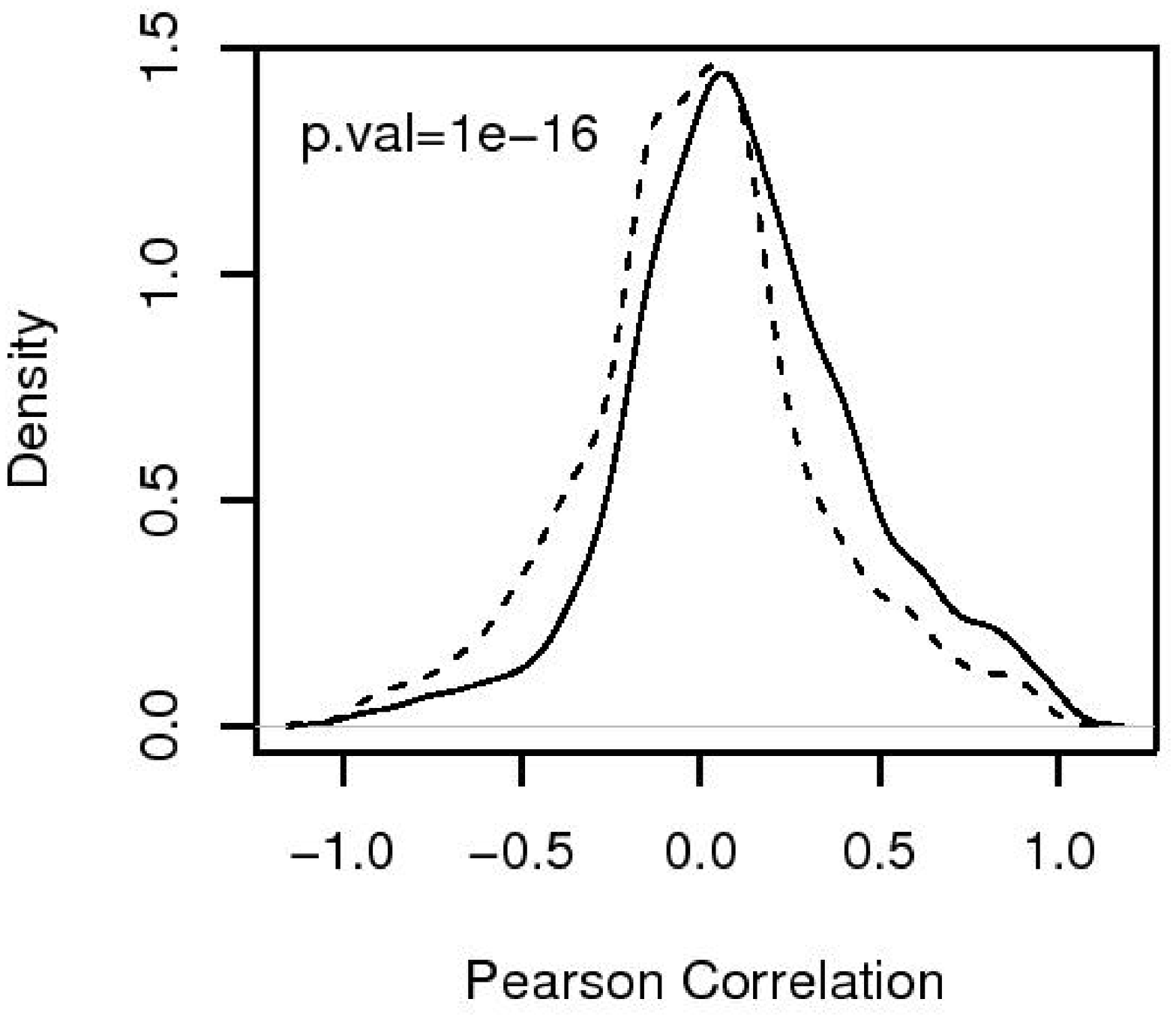}
  }
\caption{\small Left-hand plots are a comparison of the Pearson Correlation between luminosity and distance from brightest galaxy in Yang-2dF groups (top plot) and 2PIGG (bottom). In both cases the dotted line shows the expected stochastic distribution when the multiplicities are sampled according to the group occupation. Right-hand plot is the same but generated for a sample of Millennium Simulation groups.}
\label{corComp}
\end{figure*}

The results for the Pearson correlation can be found in the left hand plot of Figure \ref{corComp}, which describes the PDFs for the $P_{r}$ statistic for both 2PIGG and Yang-2dF data compared to the distribution expected for a randomly generated sample of equal multiplicity groups with randomised magnitudes and distances. It is immediately clear from these plots that whilst the 2PIGG data shows no evidence for any correlation, the Yang-2dF group data shows a strong tendency towards positive correlation. 

This difference presumably arises from the way that the groups were originally constructed. In the case of the 2PIGG FOF algorithm there is no bias towards more massive galaxies being grouped together, the linking treats all galaxies equally at a given redshift. The Yang algorithm begins with progenitor groups constructed from highly overdense regions in space or luminous single galaxies, and such a prior will clearly generate groups with dominant central galaxies. The apparent presence of positive correlation between distance to the central galaxy and the magnitude indicates that Yang-2dF groups show evidence for some sort of merger history, and thus describe a more evolved and virialised structure than 2PIGGs. This is consistent with the results from previous sections which also imply the Yang-2dF groups to be more tightly bound and virialised systems.

As a point of comparison, exactly the same methodology was applied to groups found in the Millennium Simulation \citep{spri05,de-l06}. Again the correlation was calculated between the projected distance from the central galaxy and the galaxy r-band magnitude, where the projection was chosen randomly. The results for this simulated data can be found in the right-hand plot of Figure \ref{corComp}. Immediately obvious is the extreme bias towards the positive correlation gradients expected for evolved and bound systems, with a significance even higher than for the Yang groups. KS-test statistics were calculated between all real and mock $P_{r}$ distributions and can be found in the relevant plots of Figure \ref{corComp}. Whilst the Yang-2dF groups reject a purely random distribution of galaxy magnitudes relative to the central galaxy at the 99\% level, the Millennium Simulation groups reject this at above the 7$\sigma$ level. The 2PIGG sample demonstrates an effectively random distribution of magnitudes within the groups and has a KS-test p-value of 0.44. This adds further weight to the interpretation that Yang-2dF groups describe more evolved systems, similar to those generated using semi-analytic modelling in the Millennium Simulation. De Lucia et al. (2006) interpret the simulation results as implying that the radial distance correlates with the time at which galaxies were accreted onto the central structure \citep[see also][]{gao04}.

\section{Summary}

In this paper a detailed comparison between the grouping characteristics of the 2PIGG and Yang group catalogues was made, with particular reference to the group luminosity functions and how these are affected by the different grouping algorithms. We have extended the work of Eke et al. (2004a) and Robotham et al. (2006) on the luminosity function of galaxies in groups of different masses by (a) using a maximum likelihood method for determining the LF parameters, (b) using different indicators of the group halo mass, viz. a mass direct virial mass estimate and the number of galaxies in the group.

For 2PIGG, virial masses were calculated as in Eke et al. (2004b) from the velocity dispersion and the rms distance from the group centre of the member galaxies. The same trends as previously found in \citet{robo06} are still generally present, but are somewhat less well defined for $M^*$, especially in the case of blue galaxies. The faint end slope, $\alpha$, for the total population changes in a similar way to that seen in previous work, but the subsets behave significantly differently: red galaxies now showing a large steepening, blue ones little or no change until we reach the highest masses sampled. 

In terms of ordering by multiplicity, it appears that only the groups with very few (giant) members differ substantially in $M^*$ from the larger objects. This is also the case for $\alpha$, though the blue galaxies in low multiplicity groups have the least negative slope of any blue galaxies sampled. This appears to imply that a system being numerically poor has more effect on the (lack of) blue dwarfs than the group's mass. This effect is also seen if we consider the relative numbers of galaxies brighter than and fainter than $M_{b_{j}} = -18.25$ (for groups sufficiently nearby for the latter to be visible). A surprising number of the lowest multiplicity groups have few or no galaxies in the fainter range (typically $M_{b_{j}}$ between $-18.25$ and $-17.25$).

The Yang-2dF groups provide a useful counterpoint to the 2PIGG groups as they were constructed via an algorithm which essentially works via the binding of a galaxy to a pre-existing group core. In general terms, the Yang-2dF groups demonstrate the same LF parameter trends as before. However, there are a number of distinct differences in detail.

Firstly, even for groups clearly centred at the same position in the two catalogues (i.e. representing the same halo), the LF shapes appear to diverge somewhat for the fainter galaxies, though the LF parameters are closely similar. Second, the 2PIGG groups {\em not} closely matched by a Yang-2dF group have a fainter $M^*$ and flatter $\alpha$ than the rest, whereas unmatched Yang-2dF groups have brighter $M^*$ and steeper $\alpha$, though we should also note that the LFs for these subsets are not, in fact, well fitted by a Schechter function. A two component fit would be preferred for the Yang-2dF groups.

In terms of the LF shapes, it appears that the matched and unmatched group LFs differ primarily through an excess (shortfall) of intermediate luminosity galaxies in the unmatched 2PIGG (Yang-2dF) groups. These unmatched groups are important as they are the ones which most clearly reflect the behaviour of the grouping algorithms. It appears that 2PIGG will quite generously link objects where there are numerous medium luminosity galaxies, whereas the Yang algorithm does not (presumably because they do not add sufficient mass to make the group encompass further objects). 

Physically, this indicates that groups with dominant central galaxies or dense central cores, of the sort the Yang algorithm is designed to find, are relatively deficient in medium mass objects, consistent with them being more dynamically evolved. 2PIGG, on the other hand, may tend to link subgroups, adding additional moderate mass objects which are/were the central galaxy of their own (infalling) halo.

Among the Yang-2dF groups themselves, while the errors are larger than for 2PIGG, the mass (here really just a scale version of luminosity) ordered composites show the same {\em trends} as the 2PIGG groups, with $M^*$ brightening and $\alpha$ steepening for the more luminous groups. However, as implied by the comparison above, the actual {\em values} differ, sometimes substantially. In particularly $M^*$ is $\sim 0.5$ magnitudes or more brighter at all masses for Yang-2dF groups than 2PIGG groups. Red galaxies have consistently steeper $\alpha$ in the Yang-2dF groups, while, given the errors the blue galaxies are consistent with the same slopes as those in 2PIGG (or, indeed, with no change with group mass). 

The multiplicity ordered Yang-2dF groups show the same changes relative to 2PIGG multiplicity ordered groups as seen for the mass ordered groups, and as in 2PIGG the least negative $\alpha$ is seen for red galaxies in low multiplicity groups, blue galaxies having a similar $\alpha$ at all multiplicities. Thus the Yang-2dF groups contain a larger fraction of red early type galaxies at the same group mass or multiplicity than seen in 2PIGG and this difference is largest for the highest multiplicity groups. Generally, this deficit of low-mass passive galaxies can be directly interpret as a build up of the red sequence as a function of halo mass, and indirectly as a build up of the red sequence with redshift.

LFs calculated for galaxies within 0.5 $r_{rms}$ and outside 1 $r_{rms}$ show that the outer regions have similar LF shapes for all group multiplicities for both red and blue galaxies. The central LF for late type galaxies changes slightly with multiplicity, but the major change is for centrally located early type galaxies;
$\alpha$ is significantly steeper than for the outer regions at all multiplicities and steepens even more for higher multiplicities. Looked at in a slightly different way, we have seen that the Yang-2dF groups tend to be denser and probably more evolved, and the change in population with radius for such systems is reflected in a strong rank correlation between luminosity and group-centric distance within these groups.

Thus we should finally conclude that the distribution of luminosities of galaxies in groups depends {\em both} on the mass of the group, as determined via any suitable (and consistent) manner {\em and} on the (density) structure of the group, so that different grouping algorithms will generate different luminosity distributions for the grouped galaxies.

Finally distance-rank magnitude relations were considered. Only the Yang groups demonstrated any evidence of a correlation between a galaxy's position relative to the centre and its luminosity (i.e.\ brighter galaxies have a weak tendency to be nearer the group centre). 2PIGG possessed no such gradient, the conclusion being the non-prescriptive nature of luminosity ordering in the FOF algorithm destroys any weakly observable trend. A comparison was made to the semi-analytic galaxy catalogue of De Lucia et al. (2006) built on top of the Millennium Simulation. Interestingly this theoretical work demonstrated much stronger gradients than present even in Yang-2dF groups, and may indicate a semi-analytic formalism that is too stringent.

\section*{Acknowledgments}
The authors wish to thank Bob Nichol, Gary Mamon and Richard Bower for insightful discussions regarding this work. We are pleased to thank the 2dF GRS Team
for the publicly available final data release which provided the foundation for most of this work (www2.aao.gov.au/2dFGRS). Also thanks to both the 2PIGG and Yang-2dF teams for publicly releasing their respective group catalogues. The Millennium Simulation used in section 5 was carried out by the Virgo Supercomputing Consortium at the Computing Centre of the Max-Planck Society in Garching. The semi-analytic galaxy catalogue is publicly available at www.mpa-garching.mpg.de/galform.

\newpage

\begin{table*}
 \begin{center}
 \begin{tabular}{lrrrrrrr}\hline\hline
\multicolumn{1}{l}{Environment}&
\multicolumn{1}{c}{$M^{*}$}&
\multicolumn{1}{c}{$\alpha$}&
\multicolumn{1}{c}{p-value}&
\multicolumn{1}{c}{low $M^{*}$}&
\multicolumn{1}{c}{high $M^{*}$}&
\multicolumn{1}{c}{low $\alpha$}&
\multicolumn{1}{c}{high $\alpha$}
\\ 
\hline
5-9 Blue Cen&$-19.93$&$-1.17$&$0.91$&$-20.11$&$-19.76$&$-1.27$&$-1.06$\\
5-9 Blue Out&$-19.71$&$-1.17$&$0.44$&$-19.82$&$-19.61$&$-1.22$&$-1.08$\\
5-9 Red Cen&$-19.75$&$-0.52$&$0.69$&$-19.86$&$-19.64$&$-0.62$&$-0.43$\\
5-9 Red Out&$-19.10$&$-0.21$&$0.35$&$-19.18$&$-19.03$&$-0.33$&$-0.14$\\ \hline
10-19 Blue Cen&$-20.09$&$-1.32$&$0.84$&$-20.31$&$-19.89$&$-1.43$&$-1.21$\\
10-19 Blue Out&$-19.59$&$-1.17$&$0.78$&$-19.75$&$-19.41$&$-1.28$&$-1.04$\\
10-19 Red Cen&$-20.05$&$-0.85$&$0.61$&$-20.17$&$-19.94$&$-0.93$&$-0.77$\\
10-19 Red Out&$-19.02$&$-0.35$&$0.99$&$-19.16$&$-18.89$&$-0.51$&$-0.19$\\ \hline
20+ Blue Cen&$-20.19$&$-1.41$&$0.21$&$-20.36$&$-20.03$&$-1.49$&$-1.33$\\
20+ Blue Out&$-19.59$&$-1.17$&$0.73$&$-19.81$&$-19.37$&$-1.31$&$-1.00$\\
20+ Red Cen&$-19.95$&$-1.00$&$0.01$&$-20.03$&$-19.86$&$-1.06$&$-0.95$\\
20+ Red Out&$-19.08$&$-0.21$&$0.89$&$-19.23$&$-18.94$&$-0.45$&$-0.08$\\
\hline
\end{tabular}
\caption{\small LFs for inner and outer regions of groups.}
\label{ProfLumTab}
\end{center}
\end{table*}


\begin{thebibliography}{}
\bibitem[\protect\citeauthoryear{Abell et al.}{1989}]{abel89}
Abell G.O., Corwin H.G., Olowin R.P., 1989, ApJS, 70,1
\bibitem[\protect\citeauthoryear{Baldry at al.}{2006}]{bald06} Baldry, I.K., Balogh, M.L., Bower, R.G., Glazebrook, K., Nichol, R.C., Bamford, S.P., Budavari, T., 2006, MNRAS, 2006, 373, 469
\bibitem[\protect\citeauthoryear{Balogh at al.}{2004}]{balo04} Balogh M.L. et al., 2004, MNRAS, 348, 1355
\bibitem[\protect\citeauthoryear{Bell et al.}{2004}]{bell04} Bell E.F. et al., 2004, ApJ, 608, 752
\bibitem[\protect\citeauthoryear{Bell et al.}{2006}]{bell06} Bell E.F. et al., 2006, ApJ, 640, 241
\bibitem[\protect\citeauthoryear{Benson et al.}{2003}]{bens03} Benson A.J., Bower R.G., Frenk C.S., Lacey C.G., Baugh C.M., Cole S., 2003, ApJ, 599, 38
\bibitem[\protect\citeauthoryear{Bower et al.}{2006}]{bowe06} Bower, R.G., Benson, A.J., Malbon, R., Helly, J.C., Frenk, C.S., Baugh, C.M., Cole, S., Lacey, C.G., 2006, MNRAS, 370, 645
\bibitem[\protect\citeauthoryear{Burgett et al.}{2004}]{burg04}Burgett W.S. et al., 2004, MNRAS, 352, 605
\bibitem[\protect\citeauthoryear{Carlberg et al.}{2001}]{carl01}Carlberg, R.G., Yee H.K.C., Morris, S. L., Lin, H., Hall, P.B., Patton, D.R., Sawicki, M., Shepherd, C.W., 2001, ApJ, 552, 427
\bibitem[\protect\citeauthoryear{Cattaneo et al.}{2007}]{catt07}Cattaneo A. et al., 2007, MNRAS, 377, 63
\bibitem[\protect\citeauthoryear{Cole et al.}{2000}]{cole00}Cole S., Lacey C.G., Baugh C.M., Frenk C.S., 2000, MNRAS, 319, 168
\bibitem[\protect\citeauthoryear{Cole et al.}{2005}]{cole05}Cole S. et al., 2005, MNRAS, 362, 505
\bibitem[\protect\citeauthoryear{Colless}{1989}]{coll89}Colless M.M. 1989, MNRAS, 237, 799
\bibitem[\protect\citeauthoryear{Colless et al.}{2001}]{coll01}Colless M.M. et al., 2001, MNRAS, 328, 1039
\bibitem[\protect\citeauthoryear{Collins et al.}{1995}]{coll95}Collins, C.A., Guzzo L., Nichol R.C., Lumsden S.L., 1995, MNRAS, 274, 1071
\bibitem[\protect\citeauthoryear{Conselice et al.}{2005}]{cons05}Conselice C.J., Blackburne J.A., Papovich C., 2005, ApJ, 620, 564
\bibitem[\protect\citeauthoryear{Cooray \& Milosavljevic}{2005}]{coor05}Cooray A., Milosavljevic M., 2005, ApJ, 627, L85
\bibitem[\protect\citeauthoryear{Croton et al.}{2005}]{crot05}Croton D.J. et al., 2005, MNRAS, 356, 1155
\bibitem[\protect\citeauthoryear{Cowie et al.}{2008}]{cowi08}Cowie L.L., Barger A., 2008, ApJ, 686, 72
\bibitem[\protect\citeauthoryear{Dalton et al.}{1997}]{dalt97}Dalton G., Maddox S.J., Sutherland W.J., Efstathiou G., 1997, MNRAS, 289, 263
\bibitem[\protect\citeauthoryear{De Lucia et al.}{2006}]{de-l06}De Lucia G., Springel V., White S.D.M., Croton D., Kauffmann G., 2006, MNRAS, 366, 499
\bibitem[\protect\citeauthoryear{De Lucia et al.}{2007}]{de-l07}De Lucia G. et al., 2007, MNRAS, 374, 809
\bibitem[\protect\citeauthoryear{de Vaucouleurs}{1975}]{de-v75}de Vaucouleurs G., 1975, Galaxies and the Universe, Stars and Stellar Systems Volume 9, University of Chicage Press, Chicage, IL USA, p.557
\bibitem[\protect\citeauthoryear{Dressler}{1980}]{dres80}Dressler A., 1980, ApJ, 236, 351
\bibitem[\protect\citeauthoryear{Driver et al.}{1994}]{driv94}Driver S.P., Phillipps S., Davies J.I., Morgan I., Disney M.J., 1994, MNRAS, 268, 393
\bibitem[\protect\citeauthoryear{Driver et al.}{1998}]{driv98}Driver S.P., Couch W.J., Phillipps S., 1998, MNRAS, 301, 369
\bibitem[\protect\citeauthoryear{Eke et al.}{2004a}]{eke04a}Eke V.R. et al., 2004a, MNRAS, 348, 866
\bibitem[\protect\citeauthoryear{Eke et al.}{2004b}]{eke04b}Eke V.R. et al., 2004b, MNRAS, 355, 769
\bibitem[\protect\citeauthoryear{Eke et al.}{2005}]{eke05}Eke V.R. et al., 2005, MNRAS, 362, 1233
\bibitem[\protect\citeauthoryear{Ferguson et al.}{1991}]{ferg91}Ferguson H.C., Sandage A., 1991, AJ, 101, 765
\bibitem[\protect\citeauthoryear{Gao et al.}{2004}]{gao04}Gao L., White S.D.M., Jenkins A., Stoehr F., Springel V., 2004, MNRAS, 355, 819
\bibitem[\protect\citeauthoryear{Gerke et al.}{2005}]{gerk05}Gerke B.F. et al., 2005, ApJ, 625, 6
\bibitem[\protect\citeauthoryear{Haines et al.}{2006}]{hain06}Haines C.P., La Barbera F., Mercurio A., Meduzzi P., Busarello G, 2006, ApJ, 647, L21
\bibitem[\protect\citeauthoryear{Haines et al.}{2008}]{hain08}Haines, C.P., Gargiulo, A., Merluzzi, P., 2008, MNRAS, 385, 1201
\bibitem[\protect\citeauthoryear{Hashimoto \& Oemler}{2000}]{hash00}Hashimoto Y. and Oemler A.~J., 2000, ApJ, 530, 652
\bibitem[\protect\citeauthoryear{Hogg et al.}{2003}]{hogg03}Hogg D.W. et al., 2003, ApJ, 585, L5
\bibitem[\protect\citeauthoryear{Huchra et al.}{1982}]{huch82}Huchra J.P., Geller M.J., 1982, ApJ, 257, 423
\bibitem[\protect\citeauthoryear{Kauffmann et al.}{2004}]{kauf04}Kauffmann G., White S., Heckman T., MŽnard B., Brinchmann J., Charlot S., Tremonti C., Brinkmann J., 2004, MNRAS, 353, 713
\bibitem[\protect\citeauthoryear{Lacey et al.}{1993}]{lace93}Lacey C., Cole S., 1993, MNRAS, 262, 627
\bibitem[\protect\citeauthoryear{Lewis et al.}{2002}]{lewi02}Lewis I.J., et al., 2002, MNRAS, 334, 673
\bibitem[\protect\citeauthoryear{Mamon}{2007}]{mamo07}Mamon G., 2007, Groups of Galaxies in the Nearby Universe, eds. Saviane I., Ivanov V.D., Borrissova J., Springer-Verlag, p.203
\bibitem[\protect\citeauthoryear{McLaughlin}{1999}]{mcla99} McLaughlin, D.E., 1999, AJ, 117, 2398 
\bibitem[\protect\citeauthoryear{Merchan et al.}{2002}]{merc02}Merchan M., Zandivarez A., 2002, MNRAS, 335, 216
\bibitem[\protect\citeauthoryear{Mercurio et al.}{2006}]{merc06}Mercurio, A., 2006, MNRAS, 368, 109
\bibitem[\protect\citeauthoryear{Moore et al.}{1998}]{moor98}Moore B., Lake G., Katz N., 1998, ApJ, 495, 139
\bibitem[\protect\citeauthoryear{Navarro, Frenk \& White}{1996}]{navo96}Navarro J.F., Frenk C.S., White S.D.M., 1996, ApJ, 462, 563
\bibitem[\protect\citeauthoryear{Norberg}{2002}]{norb02}Norberg, P., et al., 2002, MNRAS, 336, 907
\bibitem[\protect\citeauthoryear{Padilla}{2004}]{padi04}Padilla N., et al., 2004, MNRAS, 352, 211
\bibitem[\protect\citeauthoryear{Phillipps et al.}{1998}]{phil98}Phillipps S., Driver S.P., Couch W.J., Smith R.M., 1998, ApJ, 498, L119
\bibitem[\protect\citeauthoryear{Popesso et al.}{2005}]{pope05}Popesso P., Biviano A., Bohringer H., Romaniello M., Voges W., 2005, A\&A, 433, 415
\bibitem[\protect\citeauthoryear{Popesso et al.}{2006}]{pope06}Popesso P., Biviano A., Bohringer H., Romaniello M., 2006, A\&A, 445, 29
\bibitem[\protect\citeauthoryear{Postman et al.}{1996}]{post96}Postman M., Lubin L.M., Gunn J.E., Oke J.B., Hoessel J.G., Schneider D.P., Christensen J.A., 1996, AJ, 111, 615
\bibitem[\protect\citeauthoryear{Pracey et al.}{2005}]{prac05}Pracey M.B., Driver S.P., De Propris R., Couch W.J., Nulsen P.E.J., 2005, MNRAS, 364, 1147
\bibitem[\protect\citeauthoryear{Press, Schechter}{1974}]{pres74}Press W.H., Schechter P., 1974, ApJ, 187, 425
\bibitem[\protect\citeauthoryear{Robotham}{2008}]{robo08}Robotham A., 2008, PhD Thesis, University of Bristol
\bibitem[\protect\citeauthoryear{Robotham et al.}{2006}]{robo06}Robotham A., Wallace C., Phillipps S., De Propris R., 2006, ApJ, 652, 1077
\bibitem[\protect\citeauthoryear{Schechter}{1976}]{sche76}Schechter P.J., 1976, ApJ, 203, 297
\bibitem[\protect\citeauthoryear{Smith et al.}{2008}]{smit08}Smith R.J., et al., 2008, MNRAS, 386, L96
\bibitem[\protect\citeauthoryear{Smith et al.}{2009}]{smit09}Smith R.J., Lucey J.R., Hudson M.J., Allanson S.P., Bridges T.J., Hornschemeier A.E., Marzke R.O., Miller N.A., 2009, MNRAS, 392, 1265
\bibitem[\protect\citeauthoryear{Springel et al.}{2005}]{spri05}Springel V., et al., 2005, Nature, 435, 629
\bibitem[\protect\citeauthoryear{Stott et al.}{2007}]{stot07}Stott J.P., Smail I., Edge A.C., Ebeling H., Smith G.P., Kneib J.P., Pimbblet K.A., 2007, ApJ, 661, 95
\bibitem[\protect\citeauthoryear{Tago et al.}{2006}]{tago06}Tago E., et al., 2006, Astron. Nach., 327, 365
\bibitem[\protect\citeauthoryear{Tanaka et al.}{2005}]{tana05}Tanaka M., Kodama T., Arimoto N., Okamura S., Umetsu K., Shimasaku K., Tanaka I., Yamada T., 2005, MNRAS, 362, 268
\bibitem[\protect\citeauthoryear{Trentham et al.}{2002}]{tren02}Trentham N., Hodgkin S., 2002, MNRAS, 333, 423
\bibitem[\protect\citeauthoryear{Vale et al.}{2004}]{vale04}Vale A., Ostriker J.P., 2004, MNRAS, 353, 189
\bibitem[\protect\citeauthoryear{van den Bosch}{2002}]{van-02}van den Bosch F.C., 2002, MNRAS, 331, 98
\bibitem[\protect\citeauthoryear{van den Bosch et al.}{2003}]{van-03}van den Bosch F.C., Yang X., Mo H.J., 2003, MNRAS, 340, 771
\bibitem[\protect\citeauthoryear{Wilman et al.}{2005}]{wilm05}Wilman D.~J., at al., 2005, MNRAS, 358, 88
\bibitem[\protect\citeauthoryear{Yang et al.}{2005a}]{yang05a}Yang X., Mo H.J., Jing Y.P., van den Bosch F.C., 2005a, MNRAS, 358, 217
\bibitem[\protect\citeauthoryear{Yang et al.}{2005b}]{yang05b}Yang X., Mo H.J., van den Bosch F.C., Jing Y.P., 2005b, MNRAS, 356, 1293
\bibitem[\protect\citeauthoryear{Yang et al.}{2005c}]{yang05c}Yang X., Mo H.J., van den Bosch F.C., Jing Y.P., 2005c, MNRAS, 357, 608
\bibitem[\protect\citeauthoryear{Yang et al.}{2005d}]{yang05d}Yang X., Mo H.J., van den Bosch F.C., Jing Y.P., 2005d, MNRAS, 358, 216
\bibitem[\protect\citeauthoryear{Yang et al.}{2005e}]{yang05e}Yang X., Mo H.J., van den Bosch F.C., Weinmann A.M., Cheng L., Jing Y.P., 2005e, MNRAS, 362, 711
\bibitem[\protect\citeauthoryear{Zabludoff et al.}{2000}]{zabl00}Zabludoff A.I., Mulchaey J.S., 2000, ApJ, 539, 136
\bibitem[\protect\citeauthoryear{Zandivarez et al.}{2006}]{zand06}Zandivarez A., Martinez H.J., Merchan M.E., 2006, ApJ, 650, 137
\end{thebibliography}
\end{document}